\documentclass[journal,twocolumn,twoside]{IEEEtranTCOM}

\normalsize

\usepackage{cite}

\usepackage[cmex10]{amsmath}
\usepackage{array}

\usepackage{mdwmath}
\usepackage{mdwtab}
\usepackage{fixltx2e}
\usepackage{stfloats}
\usepackage{url}

\usepackage{comment}
\usepackage{times}
\usepackage{mathptmx}
\usepackage{newtxmath}
\usepackage[utf8]{inputenc}
\usepackage{dsfont}
\usepackage{graphicx}
\usepackage[english]{babel}

\makeatletter
\usepackage{pstricks,bbm,bm,mathrsfs,psfrag,pifont,times}
\usepackage{hyperref}
\usepackage{color}
\usepackage[normalem]{ulem}

\DeclareMathOperator*{\argmax}{arg\,max}
\def\a{ {\mathrm{I}} }
\def\b{ {\mathrm{II}} }
\def\p{ {\mathrm{p}} }

\def\receff{ \zeta}
\def\E{{\bf E} }

\makeatother

\usepackage{babel}
\begin{document}
\global\long\def\bra#1{\mathinner{\langle{#1}|}}%
\global\long\def\ket#1{\mathinner{|{#1}\rangle}}%
\global\long\def\braket#1{\mathinner{\langle{#1}\rangle}}%
\global\long\def\Id{{\mathbbm1}}%
\title{Probabilistic amplitude shaping for continuous-variable quantum
key distribution with discrete modulation over a wiretap channel}

\author{Michele~N.~Notarnicola,
        Stefano~Olivares,
	Enrico~Forestieri,~\IEEEmembership{Senior~Member,~IEEE,}
	Emanuele~Parente,
	Luca~Potì,~\IEEEmembership{Senior~Member,~IEEE,}
        and~Marco~Secondini~\IEEEmembership{Senior~Member,~IEEE} 
\thanks{Manuscript received November 10, 2022. Manuscript revised \today. This work has been partially supported by MAECI, Project No.~PGR06314 ``ENYGMA'', by University of Milan, Project No.~RV-PSR-SOE-2020-SOLIV ``S-O~PhoQuLis'', by PNRR MUR project PE0000023-NQSTI, by National Operational Programme on Research and Innovation 2014--2020 - FSE REACT EU ``Azione IV.5 Dottorati su tematiche Green'', and by Scuola Superiore Sant'Anna, Project Quantum Pathfinder.  {\em (Corresponding author: Stefano Olivares)}}
\thanks{Michele~N.~Notarnicola and Stefano~Olivares are with the Dipartimento di Fisica ``Aldo Pontremoli'', Università degli Studi di Milano, I-20133 Milano, Italy, and also with INFN, Sezione di Milano, I-20133 Milano, Italy (e-mail: stefano.olivares@fisica.unimi.it).}
\thanks{Enrico~Forestieri, Emanuele~Parente and Marco~Secondini are with the TeCIP Institute, Scuola Superiore Sant’Anna, I-56124, Pisa, Italy, and also with National Laboratory of Photonic Networks, CNIT, I-56124 Pisa, Italy (e-mail: marco.secondini@santannapisa.it).}
\thanks{Luca~Potì is with the National Laboratory of Photonic Networks, CNIT, I-56124 Pisa, Italy.}
}


\maketitle

\begin{abstract}
To achieve the maximum information transfer and face a
possible eavesdropper, the samples transmitted in continuous-variable quantum key
distribution (CV-QKD) protocols are to be drawn from a continuous Gaussian distribution.
As a matter of fact, in practical implementations the transmitter has a finite (power) dynamics and  the Gaussian sampling can be only approximated. This requires the quantum protocols to operate at small powers. In this paper, we show that a suitable probabilistic amplitude shaping of a finite set of symbols allows to approximate at will the optimal channel capacity also for increasing average powers.  
We investigate the feasibility of this approach in the framework of CV-QKD, propose a protocol employing discrete quadrature amplitude modulation assisted with probabilistic amplitude shaping, and we perform the key generation rate analysis assuming a wiretap channel and lossless homodyne detection.
\end{abstract}


\IEEEpeerreviewmaketitle

\section{Introduction}

\IEEEPARstart{C}{urrent} cryptographic systems for communications security are mostly
based on either public-key cryptography, providing only conditional security \cite{RSA:rivest1978}, or the one-time pad \cite{vernam1926cipher}, offering unconditional security (guaranteed by information theory) but with much less practical implementation \cite{Shannon49}.
Thus, a possible solution to the key distribution problem is offered by
quantum key distribution (QKD) protocols \cite{BB84:bennet1984,ekert1992quantum,grosshans2002continuous},
in which two distant parties (Alice and Bob) share a secret key by exchanging quantum states through an untrusted quantum channel under the control of an eavesdropper (Eve).

The very laws of quantum mechanics guarantee unconditional
security of QKD protocols \cite{gisin2002quantum}; any measurement
of the quantum states performed by Eve leaves a trace on the states
themselves, allowing Alice and Bob to detect the intrusion, evaluate
the amount of information possibly gained by Eve, and discard the
key if necessary \cite{wootters1982single}.

A promising approach to QKD is the one based on continuous variables
(CVs) \cite{Ralph,grosshans2002continuous,grosshans2003quantum,grosshans2005coll, grosshans2007continuous,e17096072}.
With respect to discrete-variable (DV) protocols \cite{BB84:bennet1984},
in which single photons are typically used as information carriers,
CV protocols use coherent states \cite{OLIVARES:PLA} (namely, laser pulses) to carry information, exactly as in classical communication systems. Unlike DV-QKD, CV-QKD does not
require single-photon sources and detectors and can use the same devices
and modulation/detection schemes commonly employed in classical coherent
optical communications \cite{lodewyck2005controlling,lodewyck2007quantum,fossier2009field}.
Therefore, we expect that the implementation of a CV-QKD system could
also benefit from the use of the most effective modulation/detection
techniques and digital processing strategies that have been developed
for classical systems in the last years.

Nevertheless, an important issue concerning CV-QKD is the modulation of coherent
states. Although the first CV-QKD protocols were originally based
on discrete modulations \cite{Ralph,hillery2000quantum,reid2000quantum},
in 
the protocol proposed by Grosshans and Grangier in 2002 (GG02)
the states are modulated with a Gaussian distribution
\cite{grosshans2002continuous,grosshans2003quantum, grosshans2005coll, grosshans2007continuous}. This choice is related to the fundamental objective of maximizing the key generation
rate (KGR), that is, the length of the secret key shared by Alice
and Bob per unit time 
\cite{Shannon48,Gallager68}.
Indeed, the actual KGR of a QKD system is limited
by the difference between the amount of information that can be reliably
transmitted from Alice to Bob per unit time and that obtainable by
Eve. The maximization of the first quantity (which yields the channel
capacity) is a classical problem in digital communications and is
achieved with a Gaussian input distribution.

Although justified on a theoretical level, Gaussian modulation involves
several practical difficulties and is never used in classical digital communication systems for a twofold reason \cite{Jouguet2012, DjordjevicDGM}. 
The main drawback is that any realistic transmitter may generate signals up to a 
certain maximum
peak power (e.g. by modulating the light emitted by a laser),
whereas sampling a Gaussian distribution implies a nonzero probability of drawing a very large sample, and this probability increases  with the average input power.
Hence, to avoid exceeding the
maximum peak power, the system has to be operated at small average power.
For example, if we want to generate a zero-mean Gaussian symbol $X$
with variance $\sigma^{2}$, but we are limited by hardware constraints
to the range $|X|<X_{\mathrm{max}}$, we must ensure that $X_{\mathrm{max}}>6\sigma$
to make the probability that $X$ exceeds the allowed range negligible
(lower than $2\times10^{-6}$ in this example). This means that the
average transmitted power, proportional to $\sigma^{2}$, must be
at least $36$ times ($15.6$~dB) smaller than the maximum peak power.
proportional to $X_{\mathrm{max}}^{2}$.
Furthermore, from the implementation point of view, the inevitable use of analog-to-digital converters
with finite resolution and dynamic range
at the transmitter introduces a discretization of the sampling distribution. 
Thus, a more feasible way to perform quantum communication at higher average power without exceeding the
transmitter dynamics and available resolution is that of employing discrete modulation formats of
appropriate order. 

CV-QKD employing discrete modulation has been approached with uniform phase-shift-keying (PSK) modulation \cite{leverrier2009unconditional, leverrier2010continuous, leverrier2011continuous, becir2012continuous, hirano2017implementation, Qu2017, ghorai2019asymptotic,lin2019asymptotic, Liao2020, Ghalaii2020, Papanastasiou2021, Denys2021explicitasymptotic, Kanitschar2022}, obtaining a lower KGR with respect to the GG02 scheme. However, in a practical scenario the GG02 protocol exhibits a lower reconciliation efficiency \cite{leverrier2009unconditional}, making PSK still worth of interest.
Nevertheless, for a large PSK constellation increasing further the number of symbols brings only a negligible advantage \cite{Denys2021explicitasymptotic}. To overcome these limitations, also amplitude-phase-shift-keying (APSK) formats have been investigated \cite{Djordjevic2019, Almeida2021, Pereira2022}.

More recently, quadrature-amplitude modulation (QAM) of a regular grid of signals has been proposed as a promising solution \cite{Denys2021explicitasymptotic, Roumestan2021, Roumestan2022}. Differently from PSK, QAM constellations may employ a non-uniform discrete probability distribution of the
symbols that approximates better the Gaussian one, thus obtaining a higher KGR closer to GG02. 
To implement this non-uniform sampling, probabilistic amplitude shaping (PAS) is a practical
coded modulation scheme that combines QAM, probabilistic constellation shaping, and forward error correction
(FEC) to closely approach optimal channel capacity \cite{bocherer2015bandwidth,Buchali:JLT2016,fehenberger2016probabilistic}.
PAS uses a distribution matcher to map uniformly distributed information
bits on QAM symbols with the desired target distribution \cite{CCDM:schulte2016,HIDM:civelli2020,ESS:gultekin2020}.
In particular, a Maxwell--Boltzmann target distribution is considered,
which maximizes the source entropy for a given discrete constellation
and mean energy per symbol \cite{kschischang1993optimal} (in practice,
lower-energy symbols are used more often than higher-energy symbols,
reducing the energy required to achieve a certain information rate).

PAS can be easily combined with the efficient FEC codes and the corresponding
decoding algorithms commonly employed in digital communications (e.g.,
binary LDPC codes with iterative belief propagation), obtaining an
excellent trade-off between performance and complexity \cite{buchali2015experimental,fehenberger2016sensitivity,Buchali:JLT2016}.
Ease of implementation, nearly optimal performance, fine rate tunability,
and compatibility with existing devices and techniques make PAS one
of the most popular solutions for the latest generation of coherent
optical systems. These considerations suggest that a similar solution
might be convenient also for CV-QKD. 

In this work, we consider a CV-QKD
protocol implemented thrugh QAM modulation of coherent states and homodyne detection.
We consider both uniform sampling of the coherent states and non-uniform generation assisted with PAS \cite{bocherer2015bandwidth,fehenberger2016probabilistic}.
To address the performance of the protocol, here we perform a KGR analysis for a quantum wiretap channel \cite{Cai2004, Pan2020}. In this scenario, Eve does not alter the nature of the quantum channel connecting Alice and Bob, being only limited to
collect the lost fraction of the signals exchanged and to hide behind the channel excess noise.
Although this assumption is not sufficient to prove unconditional security, it provides a realistic model and a useful benchmark for many practical scenarios 
(e.g. free space communications) \cite{Banaszek2021}. Accordingly, the paper focuses on the task of key rate optimization rather than providing a security analysis under arbitrary eavesdropping attacks.
In particular,
we consider different QAM constellations \cite{proakis01,cariolaro2015quantum}, optimize the shaping parameter and compare the performance with and without PAS with respect to a PSK modulation with the same number of symbols and the original GG02 scheme. 
At first, we address the case of a pure-loss channel and, thereafter, discuss the effects of thermal noise on the obtained results.

The structure of the paper is the following. In Sec.~\ref{sec: DiscreteProto} we present
our proposal of a CV-QKD scheme employing discrete modulation and based on
standard probabilistic shaping, namely, optimizing the discrete probability distribution over the capacity of the channel, whilst in Sec.~\ref{sec: ExNoise} we address the effect of the channel excess noise on the obtained results.
Then, in Sec.~\ref{sec: AlternativeOptimiz} we propose a new optimization procedure for PAS
and discuss its performance. Finally, in Sec.~\ref{sec: Concl} we close the paper by drawing some concluding remarks and  their future developments.

\section{CV-QKD with QAM discrete modulation and PAS \label{sec: DiscreteProto}}
\begin{figure}
\centerline{\includegraphics[width=3.2in]{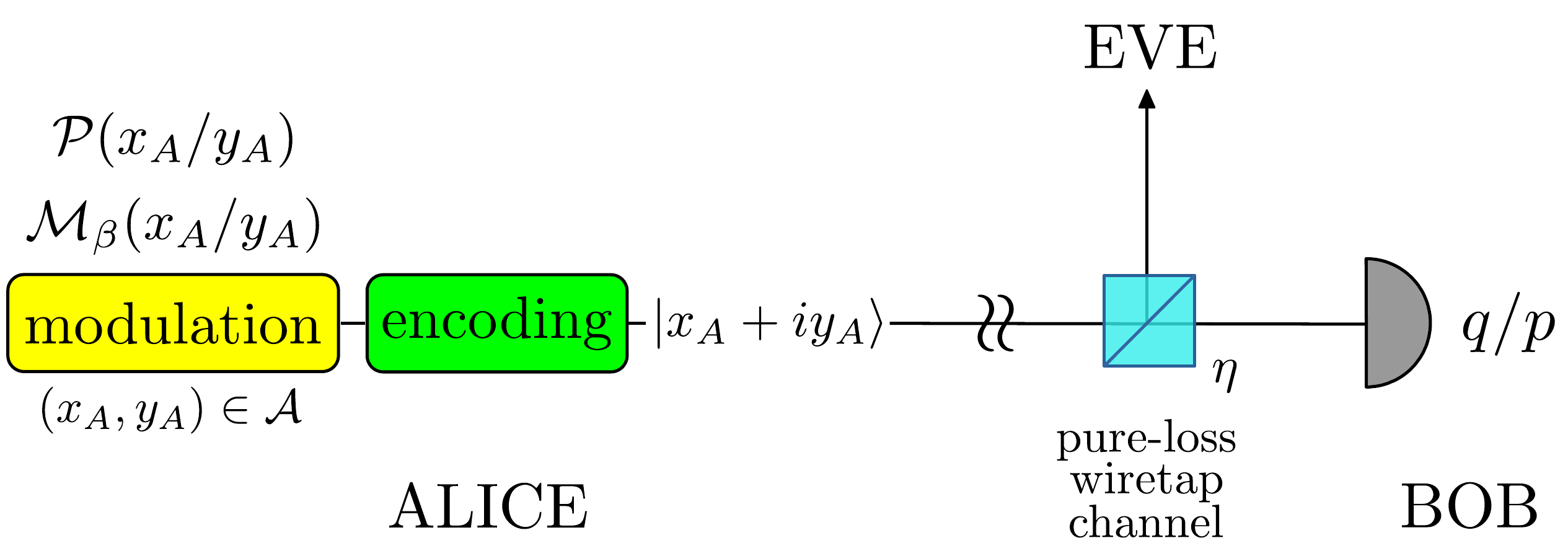} }
\caption{Schematic representation of the CV-QKD protocol discussed in the paper
for a discrete modulation format. Alice generates symbols $z=x_{A}, y_{A}$ by samping either a uniform distribution $\mathcal{P}(z)$ (case $\a$) or a Maxwell-Boltzmann distribution $\mathcal{M}_\beta(z)$ (case $\b$), encodes them onto $\protect\ket{x_{A}+iy_{A}}$ and
sends them to Bob through an untrusted pure-loss channel. Bob investigates
the channel by performing a homodyne measurement of $q/p$, chosen at random. In this scenario, Eve only collects the fraction of the signals lost during the propagation through the channel.}\label{fig:01_proto}
\end{figure}

In this paper we investigate the protocol for CV-QKD employing discrete modulation depicted in Fig.~\ref{fig:01_proto}. 
The sender, Alice ($A$), encodes the information on several laser pulses
described by coherent states $| x_A + i y_A \rangle$, where $x_A, y_A \in \mathbb{R}$ are randomly generated according to a discrete-valued probability distribution.
These states are sent to the receiver, Bob ($B$), through an untrusted quantum channel which may be attacked by an eavesdropper, Eve ($E$).

Once received the signals, Bob probes them by implementing the measurement of one of the two orthogonal quadratures $q$ or $p$, $[q,p]=2i\sigma_{0}^{2}$, where $\sigma_{0}^{2}$ represents the shot-noise variance, that is the vacuum fluctuations $\braket{0|q^{2}|0}=\braket{0|p^{2}|0}=\sigma_{0}^{2}$. The measured quadrature is chosen at random and implemented by performing homodyne detection \cite{OLIVARES:PLA}. Without loss of generality, to carry on the security analysis we can assume that Bob measures $q$.
From now on we will consider shot-noise units, that is $\sigma_0^2=1$.

Then, Alice and Bob share some correlated classical
information through a classical authenticated channel, the so-called reconciliation process:
we have either direct or reverse reconciliation (DR) if information is publicly shared by Alice or Bob, respectively \cite{grosshans2007continuous,van2004reconciliation,bennett1995generalized}. After a privacy amplification stage \cite{grosshans2007continuous}, they are finally able to generate a random secure key.

Standard optical fibers are well-modeled by thermal-loss channels \cite{lodewyck2005controlling, lodewyck2007quantum}. However, the numerical analysis of such systems is computationally demanding. Therefore, here we deal with a simpler scenario and assume a pure-loss channel, whilst the impact of a non-zero channel excess noise on the present protocol will be discussed in Sec.~\ref{sec: ExNoise}.
The pure-loss channel is described by means of a beam splitter with transmissivity $\eta= 10^{-0.1 \kappa d}$, where $d$ is the transmission distance in km, and $\kappa=0.2$ dB/km is the loss rate of common fibers \cite{agrawal02,Jouguet_Kunz-Jacques_Leverrier_Grangier_Diamanti_2013,Lodewyck_Debuisschert_Tualle-Brouri_Grangier_2005}.
In this case, it is possible to perform CV-QKD up to a large maximum transmission distance depending only on the imperfect reconciliation process \cite{leverrier2009unconditional, leverrier2010continuous, leverrier2011continuous, Denys2021explicitasymptotic}.
As a consequence, the goal of this section is to compare the QAM modulation with both the already existing PSK schemes and the Gaussian-modulated GG02 protocol in order to establish a hierarchy between these cases.
In fact, the GG02 scheme is the one that maximize the mutual information in key distribution (see App.~\ref{sec: GG02}), providing a benchmark in evaluating the performance of the presented results. On the other hand, both PSK and QAM represent sub-optimal modulation formats for quantum communications, thus their comparison is worth of interest also for CV-QKD.
In the following we will analyze in detail each step of the proposed protocol, performing a complete security analysis.

\subsection{Modulation stage}
\begin{figure}
\centerline{\includegraphics[width=3.2in]{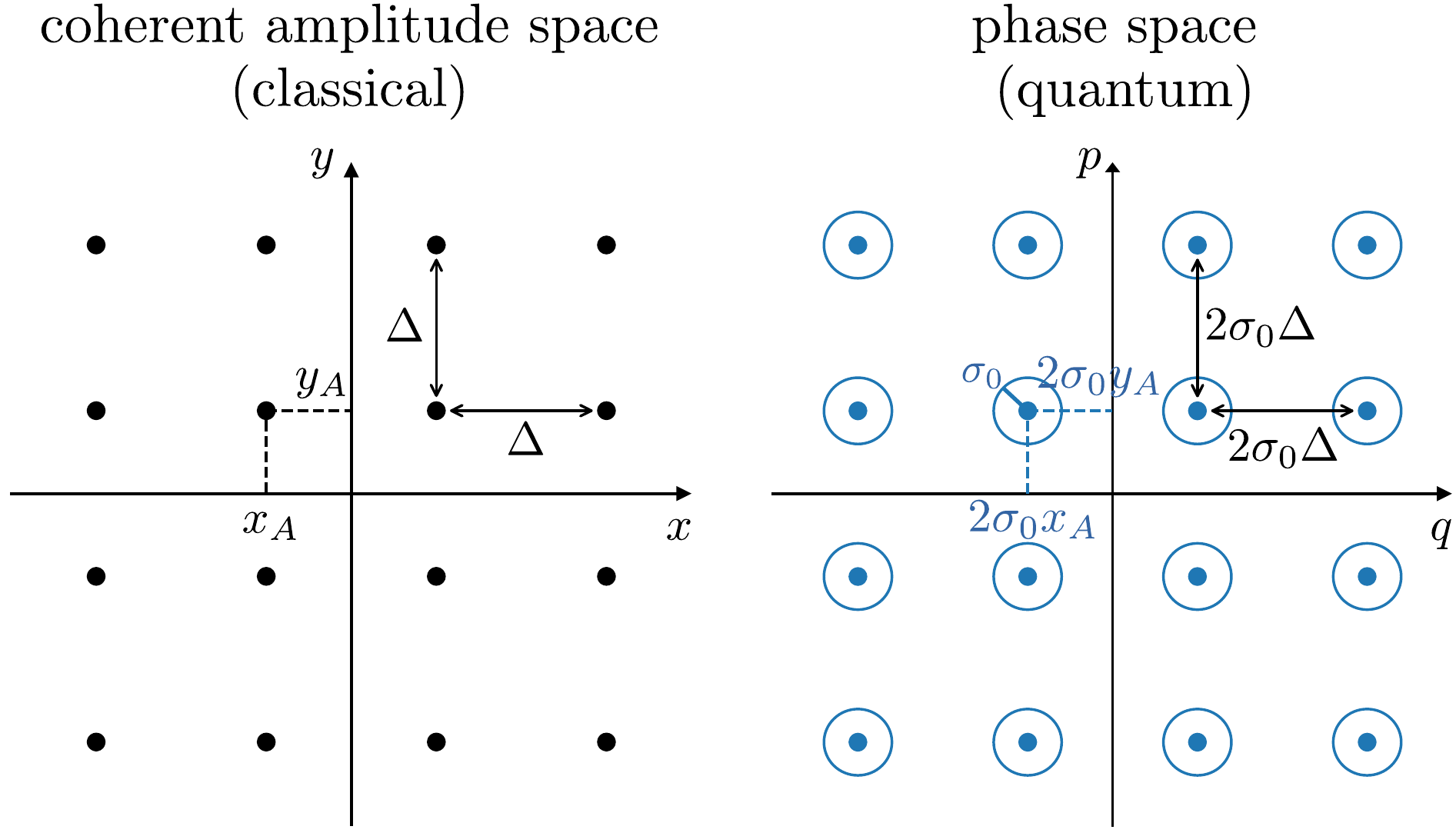} }
\caption{The QAM16 constellation ($M=4$), represented
in both the (classical) complex space of coherent amplitudes and the (quantum) phase
space.}
\label{fig:02_constellation}
\end{figure}

The QAM format adopted by Alice works as follows \cite{proakis01,cariolaro2015quantum}.
She draws each couple $(x_A, y_A)$ from the finite set
$\mathcal{A}=\Lambda\times\Lambda$,
where
$\Lambda=\{n\Delta\ |\ n=-(M-1)/2,\ldots ,(M-1)/2\}$ 
contains $M = 2^{k}$ points, $k\in\mathbb{N}$, and $\Delta \in\mathbb{R}$. Therefore, she has the constellation $\mathcal{A}$ of $M\times M$ symbols depicted in the left panel of Fig.~\ref{fig:02_constellation}, corresponding to a square lattice with pace $\Delta$. 
It is worth noting that in the standard communication notation the points in $\Lambda$ are commonly placed at distance $2\Delta$ between one another \cite{proakis01}, with $\Delta$ a scaling factor that determines the mean energy per symbol. Here, for the sake of simplicity, we adopt a different convention by letting such a distance be equal to $\Delta$,  which is hence referred to indifferently as scaling factor or symbol spacing.
Each couple $(x_A, y_A)$ is then encoded on its corresponding coherent state $\ket{x_A + i y_A}$, so that the constellation $\mathcal{A}$ is
mapped into the (quantum) phase space \cite{olivares:PhaseSpace} as a square lattice of $M\times M$ coherent states centred in $(2\sigma_{0}x_{A},2\sigma_{0}y_{A})$ and with pace $2\sigma_{0}\Delta$ (right panel of Fig.~\ref{fig:02_constellation}).

Concerning the probability distributions to sample $x_{A}$ and $y_{A}$, here we discuss two alternative possibilities.
The former, referred to as case $\a$, is the uniform distribution $\mathcal{P}(z)= M^{-1}$, $(z=x_{A},y_{A})$, commonly exploited in classical communications \cite{proakis01} and quantum-state-discrimination schemes \cite{cariolaro2015quantum}. The latter, case $\b$, is the Maxwell--Boltzmann distribution (MB) \cite{cover06,kschischang1993optimal}
\begin{eqnarray}\label{eq:MBdistr}
\mathcal{M}_{\beta}(z)=\frac{e^{-\beta z^{2}}}{Z}	\qquad(z=x_{A},y_{A})\label{eq:MB}
\end{eqnarray}
$Z=\sum_{z}e^{-\beta z^{2}}$ being the normalization constant.
Though being widely deployed in practice, the uniform distribution
is largely suboptimal, as it requires up to 1.53~dB more energy per
symbol to achieve the same mutual information as the capacity-achieving
Gaussian distribution over the 
additive white Gaussian noise (AWGN) channel \cite{kschischang1993optimal}.
On the other hand, the MB distribution is the maximum-entropy distribution
for a discrete random variable with given variance (mean energy) \cite{cover06}
and, for a sufficiently large QAM constellation, it has been shown
to closely approach channel capacity over the AWGN channel, practically
closing the 1.53~dB gap of uniform QAM constellations \cite{bocherer2015bandwidth}.

The MB distribution depends on the free parameter $\beta$ (the ``inverse temperature'') which we shall adjust properly to maximize the performance of the protocol, as discussed in the following subsection.
From a practical point of view, the i.i.d. MB-distributed
symbols required in case~II can be generated by the PAS scheme proposed
in \cite{bocherer2015bandwidth}, where a \emph{distribution
matcher} maps a sequence of uniform i.i.d. random
bits (in our case, the raw key generated for instance by a quantum
random number generator) to a sequence of QAM symbols with the desired
distribution. In particular, different distribution-matching algorithms
have been proposed, such as constant composition distribution matching
\cite{CCDM:schulte2016}, enumerative sphere shaping \cite{ESS:gultekin2020},
and hierarchical distribution matching \cite{HIDM:civelli2020}, which
can approach the ideal target distribution (i.i.d. MB symbols) with
arbitrary accuracy for sufficiently long sequences.

Though we do not claim the theoretical optimality of the proposed MB distribution
in terms of KGR, in the following we shall show numerically that it
provides a significant advantage over the uniform distribution and
closely approach the KGR of the original GG02 protocol.
In particular, we shall discuss the behavior of the present protocol when Alice has the relevant constellations QAM16 and QAM64, associated with the values $M=4,8$, respectively.

\subsection{Choice of the constellation parameters \label{sec: Optimization}}
First of all, we present the procedure employed to assign the appropriate values for $\beta$ and $\Delta$.
We fix the spacing $\Delta$ by introducing a constraint on the constellation energy. 
If the mean energy per symbol is equal to $\bar{n}$, the appropriate $\Delta$ is obtained by setting the variance of the sampling distribution equal to $\bar{n}/2$, namely
\begin{IEEEeqnarray}{CCLL}
\mathrm{Var}[z]&\equiv& \frac{1}{M}\sum_{z} z^{2}=\frac{\bar{n}}{2} &\qquad (\text{case } \a) 
\IEEEyessubnumber \label{eq: FindDeltaA} \\
\mathrm{Var}[z]&\equiv&\sum_{z}\mathcal{M}_{\beta}(z)z^{2}=\frac{\bar{n}}{2} &\qquad (\text{case } \b) \IEEEyessubnumber  \label{eq: FindDeltaB}
\end{IEEEeqnarray}
such that the overall state by Alice is described by the density matrix
\begin{IEEEeqnarray}{CCL}\label{eq: rho_A}
\rho_{A}^{(\a)}&=&\frac{1}{M^2}\sum_{x_A, y_A}\ket{x_{A}+iy_{A}}\bra{x_{A}+iy_{A}}
\IEEEyessubnumber \\
\rho_{A}^{(\b)}&=&\sum_{x_A,y_A}\mathcal{M}_{\beta}(x_{A})\mathcal{M}_{\beta}(y_{A})\ket{x_{A}+iy_{A}}\bra{x_{A}+iy_{A}} \IEEEyessubnumber
\end{IEEEeqnarray}
having mean energy $\bar{n}$.
For case $\a$, Eq.~(\ref{eq: FindDeltaA}) leads to the solution:
\begin{eqnarray}
\Delta^{(\a)}= \sqrt{\frac{6 \bar{n}}{M^2-1}} \, .
\end{eqnarray}
The corresponding mutual information between Alice and Bob reads
\begin{eqnarray}\label{eq: IABa}
I_{AB}^{(\a)} =H_{B}^{(\a)}- \frac{1}{M}\sum_{x_A} H_{B | x_A}	
\end{eqnarray}
where $H_{B | x_A}$ and $H^{(\a)}_{B}$ are the Shannon entropies associated with Bob's conditional probability distribution
\begin{eqnarray}
p_{B|A}(x_{B}\mid x_{A})=\frac{\exp\big[-(x_{B}-2\sigma_{0}\sqrt{\eta}x_{A})^{2}/2\sigma_{0}^{2}\big]}{\sqrt{2\pi\sigma_{0}^{2}}}	\label{eq: pB|A}
\end{eqnarray}
namely,
\begin{eqnarray}
H_{B | x_A} = \frac12 \log_2 \big(2\pi e \sigma_0^2\big)
\end{eqnarray}
and Bob's average probability distribution
\begin{eqnarray}\label{eq: pBa}
p_{B}^{(\a)}(x_{B})=\frac{1}{M} \sum_{x_A} p_{B|A}(x_{B}\mid x_{A}) \, .
\end{eqnarray}
The Shannon entropy of $p_{B}^{(\a)}(x_{B})$ has to be computed via numeric integration. In particular, in our calculations we exploited the Simpson's rule \cite{GoldenSection}.

On the contrary, Eq.~(\ref{eq: FindDeltaB}) can be handled numerically for case $\b$  and the corresponding numeric solution $\Delta^{(\b)}(\beta)$ exhibits an implicit dependence on the free parameter $\beta$.
Therefore, in this scenario we decided to optimize $\beta$ to achieve the maximum mutual information between Alice and Bob
\begin{eqnarray}\label{eq: IABb}
I_{AB}^{(\b)}(\beta)=H_{B}^{(\b)}- \sum_{x_A} \mathcal{M}_\beta (x_A) H_{B | x_A}	
\end{eqnarray}
where  $H_{B}^{(\b)}$ is the Shannon entropy associated with the distribution
\begin{eqnarray}\label{eq: pBb}
p_{B}^{(\b)}(x_{B})=\sum_{x_A} \mathcal{M}_\beta (x_A) p_{B|A}(x_{B}\mid x_{A}) \, ,
\end{eqnarray}
to be computed numerically with the Simpson's rule \cite{GoldenSection}.
Thereafter, the maximization of $I_{AB}^{(\b)}(\beta)$ has been performed with a golden-section
search algorithm \cite{GoldenSection}.
This procedure leads to the optimized inverse temperature $\beta^{(\b)}$ and, consequently, its associated spacing $\Delta^{(\b)}$, together with the optimized mutual information 
\begin{eqnarray}
I_{AB}^{(\b)}=I_{AB}^{(\b)}\left(\beta^{(\b)}\right)\,.\label{eq: opt IAB}
\end{eqnarray}

\begin{figure}
\centerline{\includegraphics[width=3.0in]{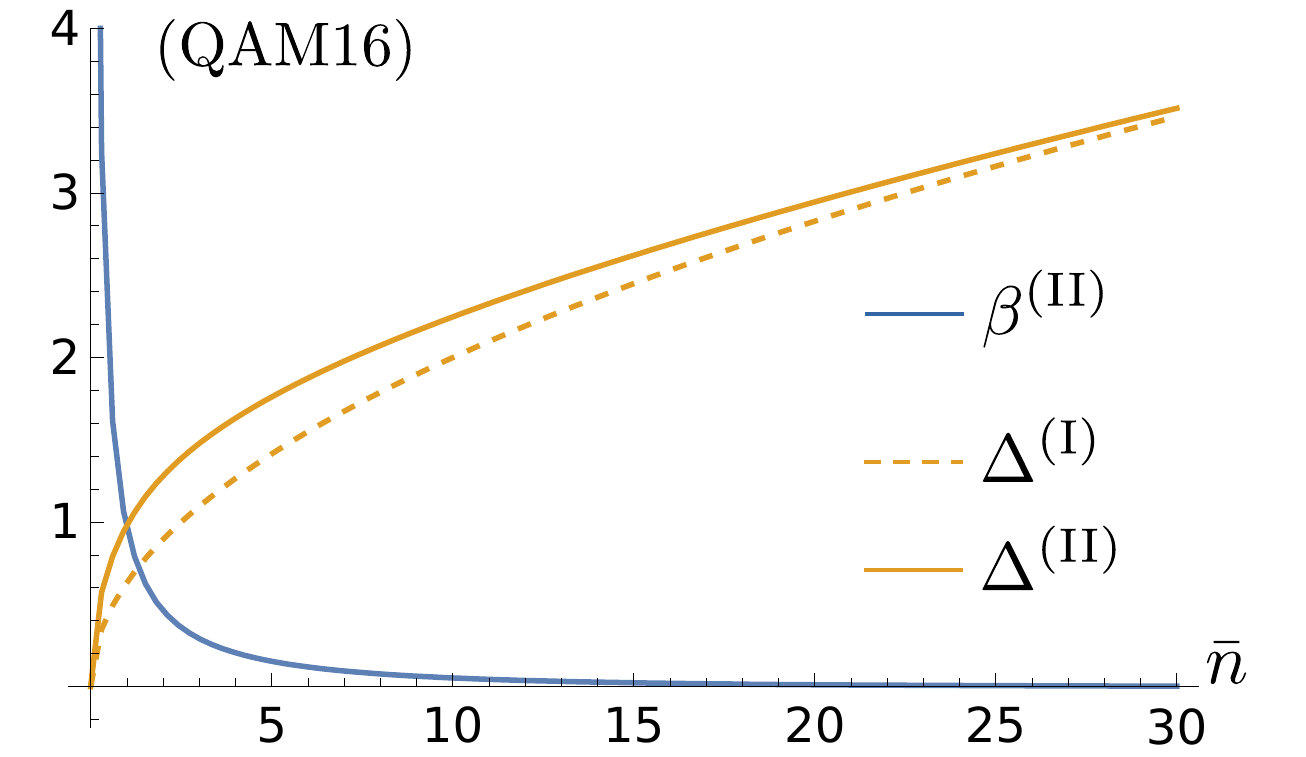} }\vspace*{2ex}
\centerline{\includegraphics[width=3.0in]{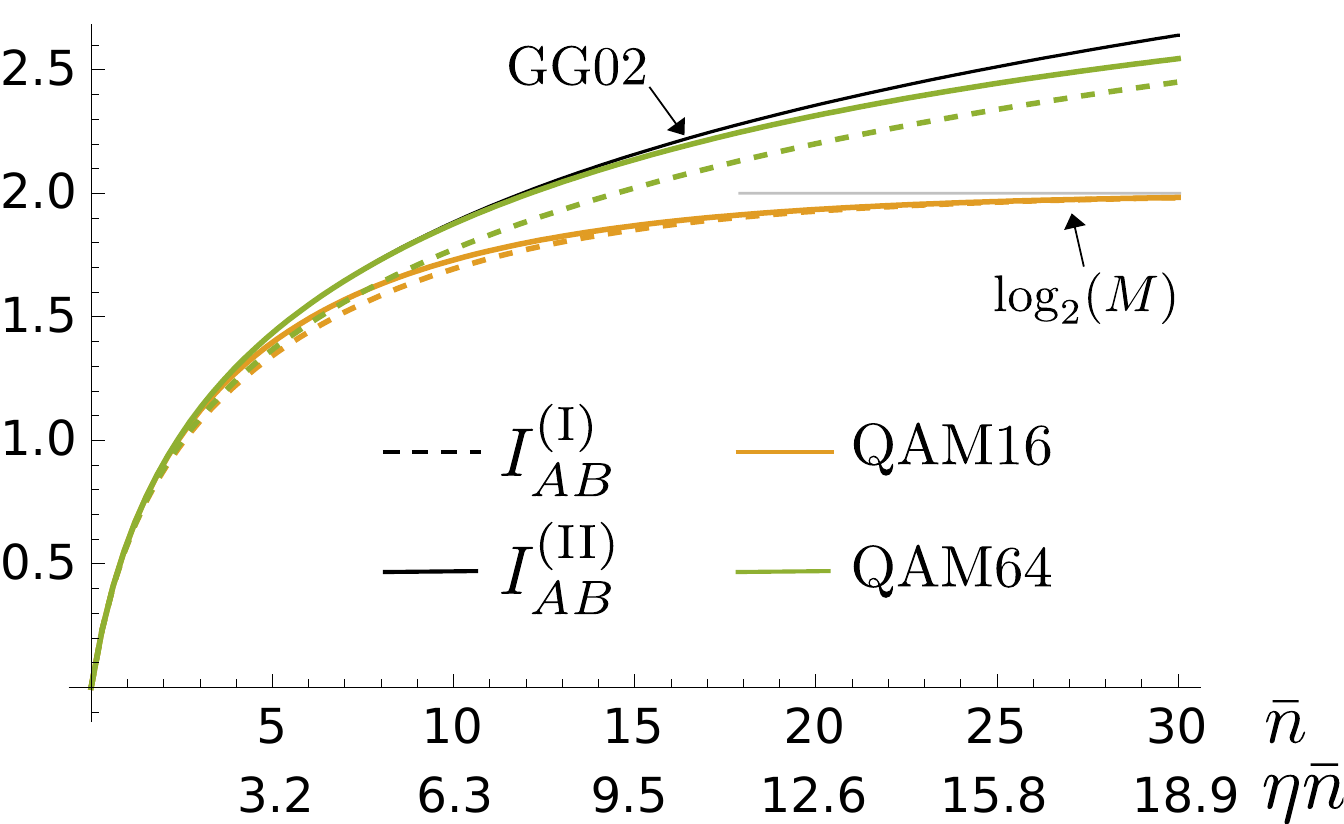} }
\caption{(Top) Optimal inverse temperature $\beta^{(\b)}$ and spacing $\Delta^{(\p)}$, $\p=\a,\b$, 
as a function of the mean energy per symbol $\bar{n}$ for a QAM16 constellation ($M=4$).
(Bottom) Optimized mutual information $I_{AB}^{(\p)}$
as a function of the modulation energy $\bar{n}$. For a better clarity we reported on the x-axis also the energy of Bob's signals $\eta \bar{n}$. The black line is the capacity of the channel, achieved with Gaussian modulation (see App.~\ref{sec: GG02}). In both the plots we fixed $d=10$ km ($2$~dB attenuation).}
\label{fig:03_optimization}
\end{figure}

The numerical results are shown in Fig.~\ref{fig:03_optimization} (top panel)
for the case of a QAM16 constellation as a function of the mean energy per symbol $\bar{n}$ and at fixed transmission distance $d=10$ km. The behavior is expected to be qualitatively equivalent for all distances.
The optimal inverse temperature $\beta^{(\b)}$ is a decreasing function of $\bar{n}$.
As we can see, for small values of $\bar{n}$, only the lowest-energy level 
of the MB in ~(\ref{eq:MB}) have non-zero probability, $\beta^{(\b)}\to \infty$,
so that the optimal constellation tends to a simple QAM4, i.e. with $M=2$;
on the other hand, for large $\bar{n}$, all the levels of the MB distribution
have the same probability, $\beta^{(\b)} \rightarrow 0$,
so that the optimal constellation tends to a uniform QAM16.
Accordingly, the non-uniform sampling of the symbols makes the spacing $\Delta^{(\b)}$ an increasing function of the energy such that $\Delta^{(\b)}\geq \Delta^{(\a)}$.

The corresponding mutual information, plotted in the bottom panel of Fig.~\ref{fig:03_optimization}, shows that the MB sampling increases the information shared between Alice and Bob, as $I_{AB}^{(\b)}\geq I_{AB}^{(\a)}$.
Moreover, in accordance with the previous considerations,
for $\bar{n} \ll 1$ all the QAM modulation formats are able to reach the capacity of the channel, achieved with Gaussian modulation (see App.~\ref{sec: GG02}); the higher $M$, the larger the region where the QAM format performs optimally. Instead,
in the large energy regime the sampling distribution becomes more and
more uniform and all the curves saturate to the maximum possible entropy of the constellation, namely 
$\log_{2} (M)$.

\subsection{KGR analysis}\label{sec: KGR}

Given the parameters $\beta$ and $\Delta$, we perform the KGR analysis for the  quantum wiretap channel \cite{Cai2004, Pan2020, Banaszek2021}. In particular, we focus on the relevant case of reverse reconciliation.
We remark that the present analysis is not sufficient to guarantee unconditional security, as Eve in principle may attack the channel and insert some suitable non-Gaussian features without being intercepted by Alice and Bob. A complete security analysis for this protocol may be carried on by exploiting the optimality of Gaussian attacks \cite{GaussOpt1, GaussOpt2, GaussOpt3, Laudenbach_Rev, Denys2021explicitasymptotic} and will be the object of future publications. 
Nevertheless, wiretap channels are gaining interest as they represent a realistic eavesdropping model, being feasible with the current technologies \cite{Pan2020, Banaszek2021}.

Moreover, we note that the uniform distribution may be retrieved from Eq.~(\ref{eq:MBdistr}) by fixing $\beta=0$. Thus, for the sake of simplicity in the following we will perform the entire analysis by considering the sole MB distribution, assuming that cases $\a$ and $\b$ correspond to $\beta=0$ and $\beta=\beta^{(\b)}$, respectively. 

As depicted in Fig.~\ref{fig:01_proto}, the pure-loss quantum wiretap channel is modeled as a beam splitter of transmissivity $\eta$ in which Eve has only access to its second port, intercepting the reflected fraction $1-\eta$ of Alice's signal. In turn, if Alice sends the state $|x_A+i y_A\rangle$, Eve receives the state $|\sqrt{1-\eta}(x_A+i y_A)\rangle$, whereas the transmitted pulse $|\sqrt{\eta}(x_A+i y_A)\rangle$ reaches Bob.
The KGR can be computed as
\begin{eqnarray}
K^{(\p)}=\receff  I_{AB}^{(\p)}-\chi_{BE}^{(\p)} \qquad (\p=\a,\b) 	\label{eq: Krcoll}
\end{eqnarray}
where $\receff \le 1$ is the reconciliation efficiency \cite{leverrier2011continuous,ghorai2019asymptotic} and
$\chi_{BE}^{(\p)}$ is the Holevo information shared between Bob and Eve \cite{holevo1998capacity}.
We recast the problem in the prepare-and-measure picture \cite{grosshans2002continuous,grosshans2003quantum,grosshans2005coll, Laudenbach_Rev}, where the Holevo information reads
\begin{eqnarray}\label{eq: chiBEp}
\chi_{BE}^{(\p)} =S\big[\rho_{E}^{(\p)}\big]-\int dx_{B} \, p_B^{(\p)} (x_{B}) \, S\big[\rho_{E|x_{B}}^{(\p)} \big] 
\end{eqnarray}
where $\rho_{E}^{(\p)}$ is the overall Eve's state, $p_B^{(\p)} (x_{B})$ is Bob's probability distribution in Eq.s~(\ref{eq: pBa}) and~(\ref{eq: pBb}), $\rho_{E|x_{B}}^{(\p)}$ is Eve's conditional state after Bob's measurement, and $S[\rho]=-\mathrm{Tr}(\rho \log_2 \rho)$ denotes the von Neumann entropy associated with the density matrix $\rho$ \cite{cariolaro2015quantum}.

To calculate these quantum states we start from the joint state of Bob and Eve,
$\rho_{BE}^{(\p)}=U_{\mathrm{BS}}(\eta)\rho_{A}^{(\p)}\otimes\ket{0}\bra{0}U_{\mathrm{BS}}(\eta)^{\dagger}$,
with $\rho_{A}^{(\p)}$ given in Eq.~(\ref{eq: rho_A}) and $U_{\mathrm{BS}}$ being the unitary operator associated with the beam splitter with transmissivity $\eta$ \cite{OLIVARES:PLA} depicted in Fig.~\ref{fig:01_proto}. After straightforward calculation we get
\begin{IEEEeqnarray}{CCL}
\IEEEyesnumber
\rho_{BE}^{(\p)} &=& \sum_{x_{A},y_{A}}\mathcal{M}_\beta(x_{A})\mathcal{M}_\beta(y_{A}) \ket{\sqrt{\eta}(x_{A}+iy_{A})}\bra{\sqrt{\eta}(x_{A}+iy_{A})}
\notag\\
& & \hspace{1.0 cm} 
 \otimes\ket{\sqrt{1-\eta}(x_{A}+iy_{A})}\bra{\sqrt{1-\eta}(x_{A}+iy_{A})}\,.
\end{IEEEeqnarray}
Then, Eve's state reads \cite{olivares:PhaseSpace}
\begin{IEEEeqnarray}{CCL}
\IEEEyesnumber
\label{eq: rhoE}
\rho_{E}^{(\p)} &=& \sum_{x_{A},y_{A}}\mathcal{M}_\beta(x_{A})\mathcal{M}_\beta(y_{A}) \notag\\
& & \hspace{0.5 cm}  \times
\ket{\sqrt{1-\eta}(x_{A}+iy_{A})}\bra{\sqrt{1-\eta}(x_{A}+iy_{A})}
\end{IEEEeqnarray}
and, if Bob gets the outcome $x_{B}$ Eve's conditional state writes \cite{olivares:PhaseSpace}
\begin{IEEEeqnarray}{CCL}
\IEEEyesnumber \label{eq: rho_E|b}
\rho_{E|x_{B}}^{(\p)}&= \frac{1}{p_{B}^{(\p)}(x_{B})}\sum_{x_{A},y_{A}} \mathcal{M}_\beta (x_{A}) \mathcal{M}_\beta (y_{A}) p_{B|A}(x_B \mid x_A) 
\nonumber\\[1ex]
& \hspace{0.5cm} \times
\, \ket{\sqrt{1-\eta}(x_{A}+iy_{A})}\bra{\sqrt{1-\eta}(x_{A}+iy_{A})} \, .
\end{IEEEeqnarray}

Calculations of the Von Neumann entropy require a numerical diagonalization of $\rho_{E}^{(\p)}$
and $\rho_{E|x_{B}}^{(\p)}$. 

\begin{figure}
\centerline{\includegraphics[width=3.0in]{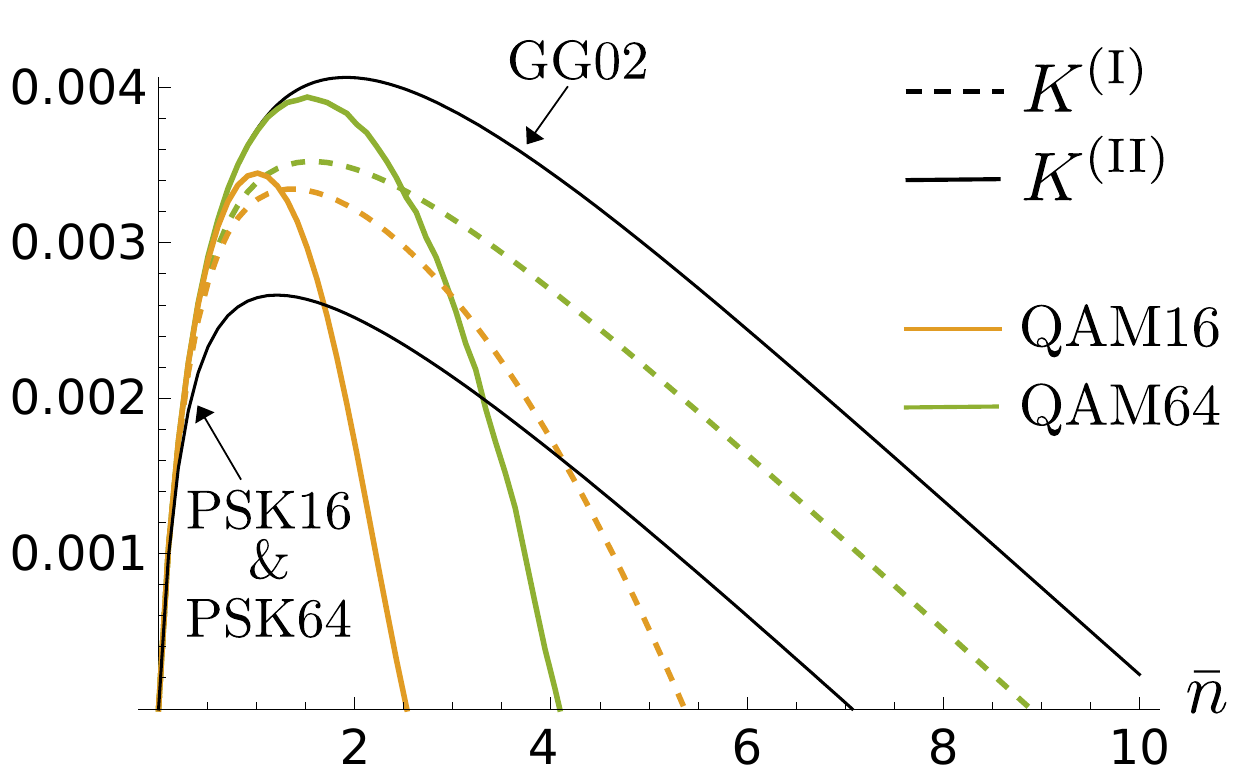} }
\caption{Plot of $K^{(\p)}$, $\p=\a,\b$, as a function of the mean energy $\bar{n}$ for $d=100$ km  ($20$~dB attenuation). The reconciliation efficiency is $\receff =0.95$.}
\label{fig:04_KGRcoll-En}
\end{figure}

The resulting $K^{(\p)}$ is plotted in Fig.~\ref{fig:04_KGRcoll-En} as a function of the mean energy $\bar{n}$ at a fixed transmission distance $d=100$ km. As expected for CV-QKD with imperfect reconciliation, all the curves reach a maximum $K^{(\p)}_{\rm max}$ for a finite value of $\bar{n}$ and ultimately become negative \cite{leverrier2009unconditional, leverrier2010continuous, leverrier2011continuous, Denys2021explicitasymptotic}. Remarkably, the MB sampling increases the maximum value of KGR with respect to the uniform one, but the curves decrease faster towards negative values. This behaviour may be interpreted as follows. For large $\bar{n}$, the larger spacing between symbols $\Delta^{(\b)} \ge \Delta^{(\a)}$ makes them more ``distinguishable", allowing Eve to retrieve more information with respect to case $\a$. On the other hand, if $\bar{n}$ is small, the shot noise becomes relevant increasing the overlap among the encoded symbols and the larger spacing is beneficial to achieve a better approximation of the Gaussian modulation.

Moreover, in Fig.~\ref{fig:04_KGRcoll-En} we also compare the performance of the QAM formats with two relevant benchmarks: the corresponding PSK constellations having the same number of symbols \cite{cariolaro2015quantum,leverrier2009unconditional,leverrier2010continuous,leverrier2011continuous} and the GG02 protocol discussed in App.~\ref{sec: GG02}.
The PSK constellation with $N$ symbols is constructed by generating uniformly the coherent states $\{|\alpha \exp[i(2k+1)\pi/N)] \rangle \}_{k=0,\ldots, N-1}$, and $\alpha=\sqrt{\bar{n}}>0$ \cite{cariolaro2015quantum}. However, by performing the above analysis with PSK, the numerical results show that increasing the size of the constellation brings a negligible increase in the KGR for large $N$  \cite{Denys2021explicitasymptotic}, and, for PSK16 and PSK64, they are almost indistinguishable.
On the contrary, in the presence of QAM, increasing the number of symbols allows to improve further the KGR, thus approaching incrementally the GG02 protocol.
Thus, in principle it is always possible to find a finite $M\times M$ QAM alphabet, with sufficiently large $M$ and optimized MB distribution, to approach with arbitrary accuracy the GG02.

\begin{figure}
\centerline{\includegraphics[width=3.0in]{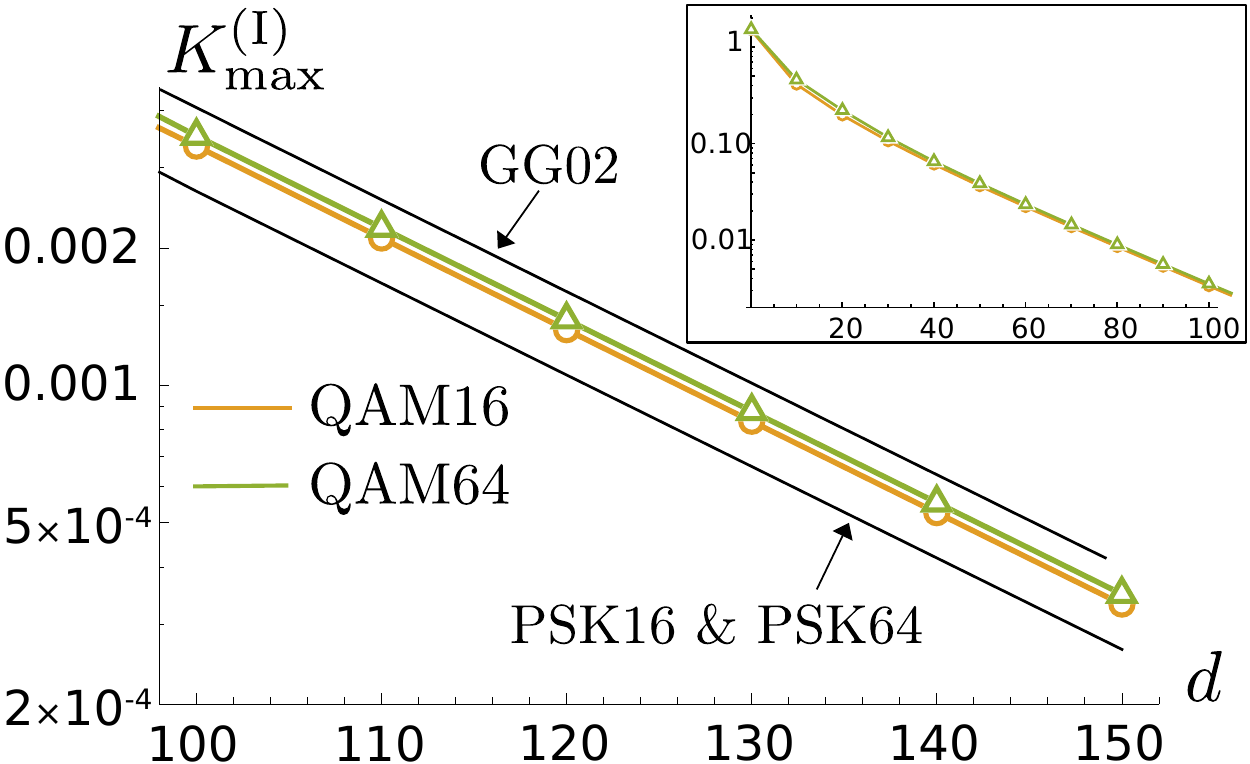} }\vspace*{3ex}
\centerline{\includegraphics[width=3.0in]{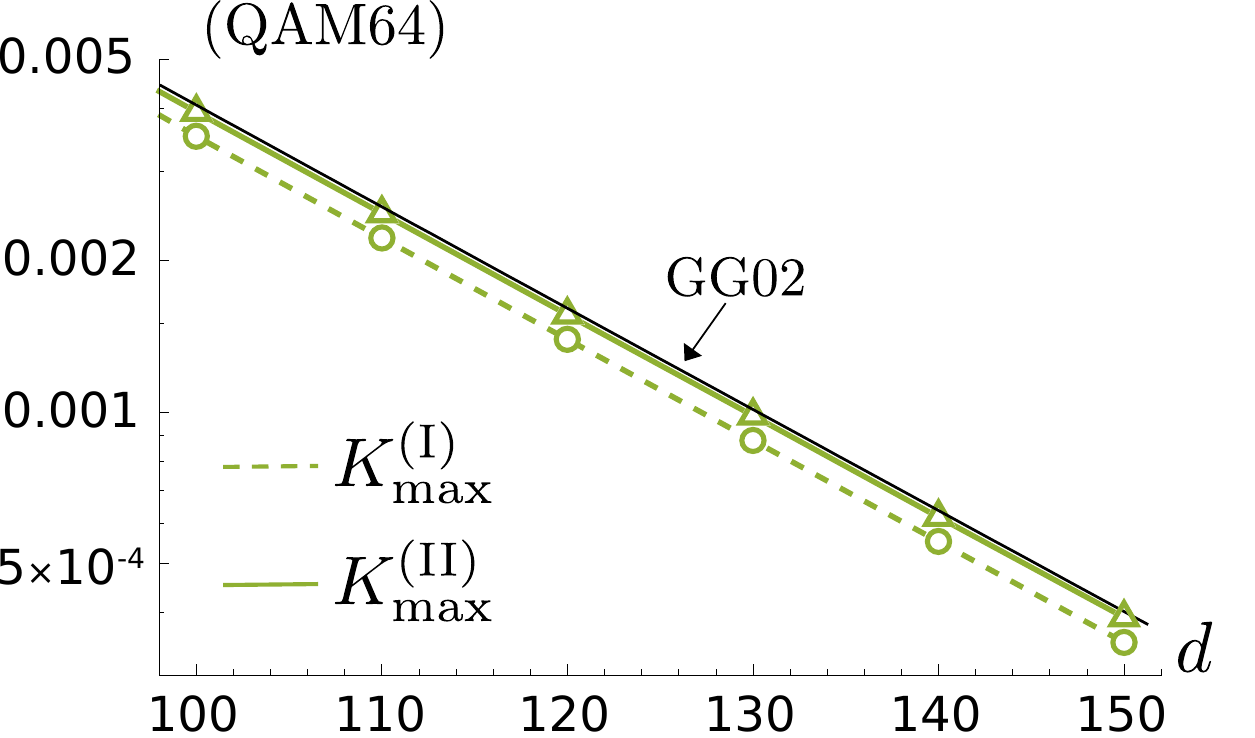}}
\caption{(Top) Log plot of $K^{(\a)}_{\rm max}$ as a function of the transmission distance $d$, expressed in km, for $d\ge 100$ km. The plot for smaller $d$ is depicted in the inset. (Bottom) Log plot of $K^{(\a)}_{\rm max}$ and $K^{(\b)}_{\rm max}$ for QAM64 ($M=8$) as a function of the transmission distance $d$. The exploitation of PAS increases the KGR, coming closer to the GG02 protocol. The reconciliation efficiency is $\receff =0.95$. The symbols (circles and triangles) refer to the numerical calculations at fixed $d$ and the lines to the corresponding interpolations.}
\label{fig:05_KGRcoll-Dist}
\end{figure}

The previous analysis may be performed to retrieve the maximum achievable KGR $K^{(\p)}_{\rm max}$, $\p=\a,\b$, as a function of the transmission distance $d$, together with the associated maximum modulation energy $\bar{n}_{\rm max}^{(\p)}$.
Plots of $K^{(\a)}_{\rm max}$ are depicted in Fig.~\ref{fig:05_KGRcoll-Dist} (top panel). 
As we can see, the QAM formats outperform the corresponding PSK ones and are closer to the GG02 protocol.
Furthermore, exploiting PAS allows to increase further the performance, as shown in the bottom panel of Fig.~\ref{fig:05_KGRcoll-Dist} for the case QAM64.

\begin{figure}
\centerline{\includegraphics[width=3.0in]{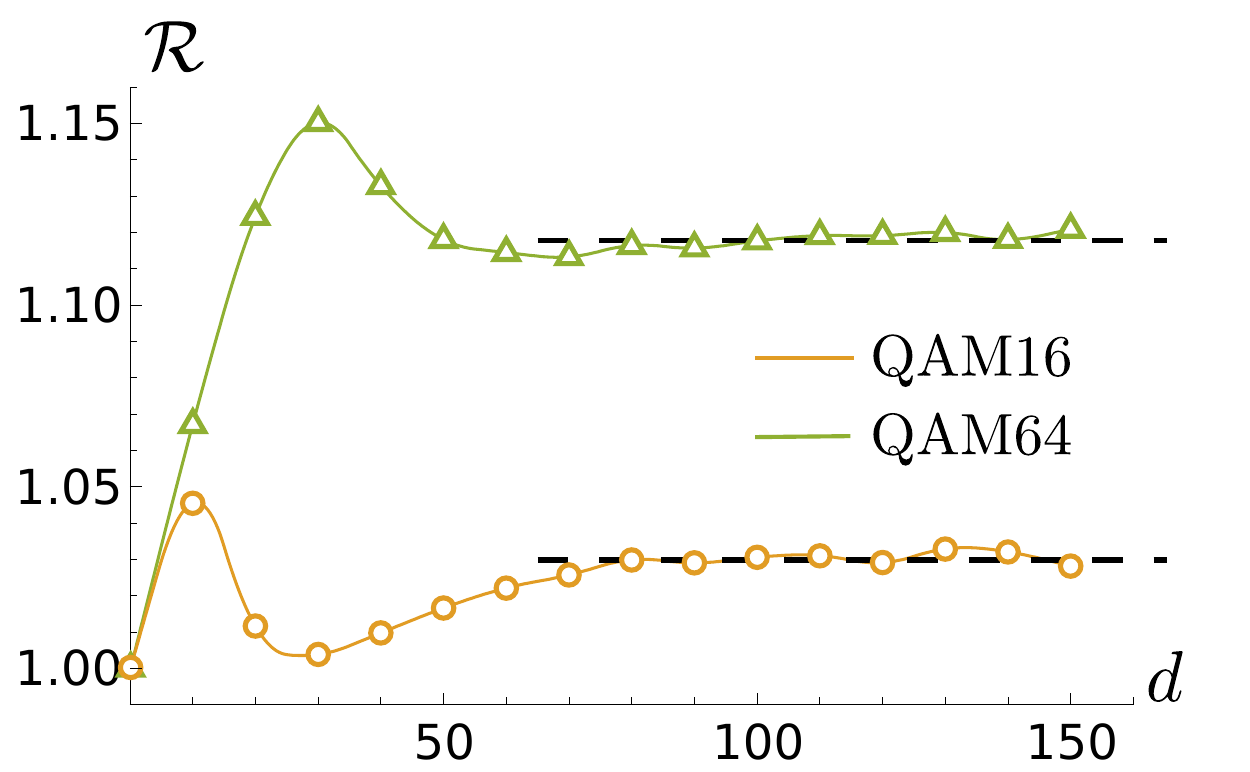}  }\vspace*{3ex}
\centerline{\includegraphics[width=3.0in]{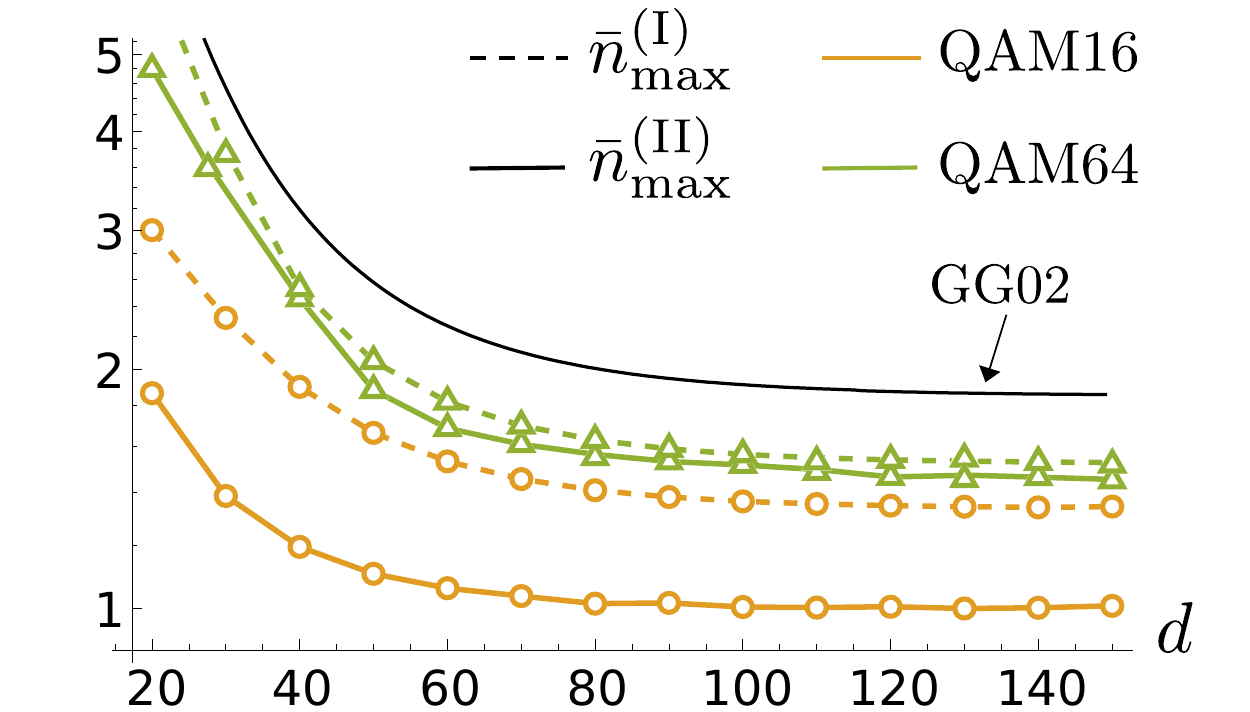} }
\caption{(Top) Plot of the ratio $\mathcal{R}$ as a function of the transmission distance $d$, expressed in km. The black dotted lines are the average of the values obtained for $d\ge 80$ km. For large distances PAS introduces an increase in the KGR equal to $\approx 3 \%$ and $\approx 12 \%$ for QAM16 and QAM64, respectively. (Bottom) Log plot of the maximum modulation energy $\bar{n}_{\rm max}^{(\p)}$, $\p=\a,\b$, as a function of the transmission distance $d$. The reconciliation efficiency is $\receff =0.95$.}
\label{fig:06_Ratio}
\end{figure}

To quantify the improvement introduced by PAS, we compute the ratio
\begin{eqnarray}
\mathcal{R}= \frac{K^{(\b)}_{\rm max}}{K^{(\a)}_{\rm max}}
\end{eqnarray}
plotted in the top panel of Fig.~\ref{fig:06_Ratio}. The trend is similar for both QAM16 and QAM64. For small $d$, $\mathcal{R}$ exhibits a bump after which it decreases until to reach an asymptotic value for $d\ge 80$ km.
We note that in this asymptotic regime the results obtained show numerical fluctuations. These are induced by a limitation of the optimization procedure described in Sec.~\ref{sec: Optimization}. Indeed, for small transmissivity $\eta \ll 1$, the overall state received by Bob has a so weak average energy $\eta \bar{n} \ll 1$, that the mutual information for cases $\a$ and $\b$ is nearly identical to the channel capacity (see Fig.~\ref{fig:03_optimization}, bottom panel). In turn, $I_{AB}^{(\b)}(\beta)$ is almost independent of $\beta$ and the maximization algorithm used in Sec.~\ref{sec: Optimization} introduces numerical fluctuations on the optimized inverse temperature $\beta^{(\b)}$.
Although these fluctuations do not affect the mutual information itself, they do have a relevant effect when computing the KGR, as appears in the plots of Fig.~\ref{fig:06_Ratio}.
To avoid this intrinsic limitation, we estimate the asymptotic ratio by calculating the average of the data obtained for $d\ge 80$ km, bringing us to an increase of $\approx 3 \%$ and $\approx 12 \%$ for QAM16 and QAM64, respectively.

Finally, for the sake of completeness, the bottom panel of Fig.~\ref{fig:06_Ratio} reports the maximum modulation energy $\bar{n}_{\rm max}^{(\p)}$, $\p=\a,\b$, which is a decreasing function of the distance and saturates as $d$ increases. As one may expect, increasing the size of the constellation increases also $\bar{n}_{\rm max}^{(\p)}$, until to reach the value of the GG02 scheme \cite{Denys2021explicitasymptotic}. Moreover, we have $\bar{n}_{\rm max}^{(\b)} \le \bar{n}_{\rm max}^{(\a)}$, that is the MB reaches its maximum KGR for lower energies, in accordance with Fig.~\ref{fig:04_KGRcoll-En}.

As
a final remark, we note that the KGR analysis performed here does
not focus on the actual error correction code and algorithm employed
for information reconciliation, but simply uses the reconciliation
efficiency parameter $\receff$ \cite{leverrier2011continuous,ghorai2019asymptotic}
to account for deviations from the KGR achievable with ideal reconciliation.
The study of ad hoc information reconciliation protocols, the impact
of possible error propagation when inverting the mapping performed
by the distribution matcher, and the comparison between the complexity
and efficiency of the information reconciliation protocols available
for Gaussian modulations and for the proposed discrete modulation
are left for future study.

\section{Role of the channel excess noise}\label{sec: ExNoise}
\begin{figure}
\centerline{\includegraphics[width=3.2in]{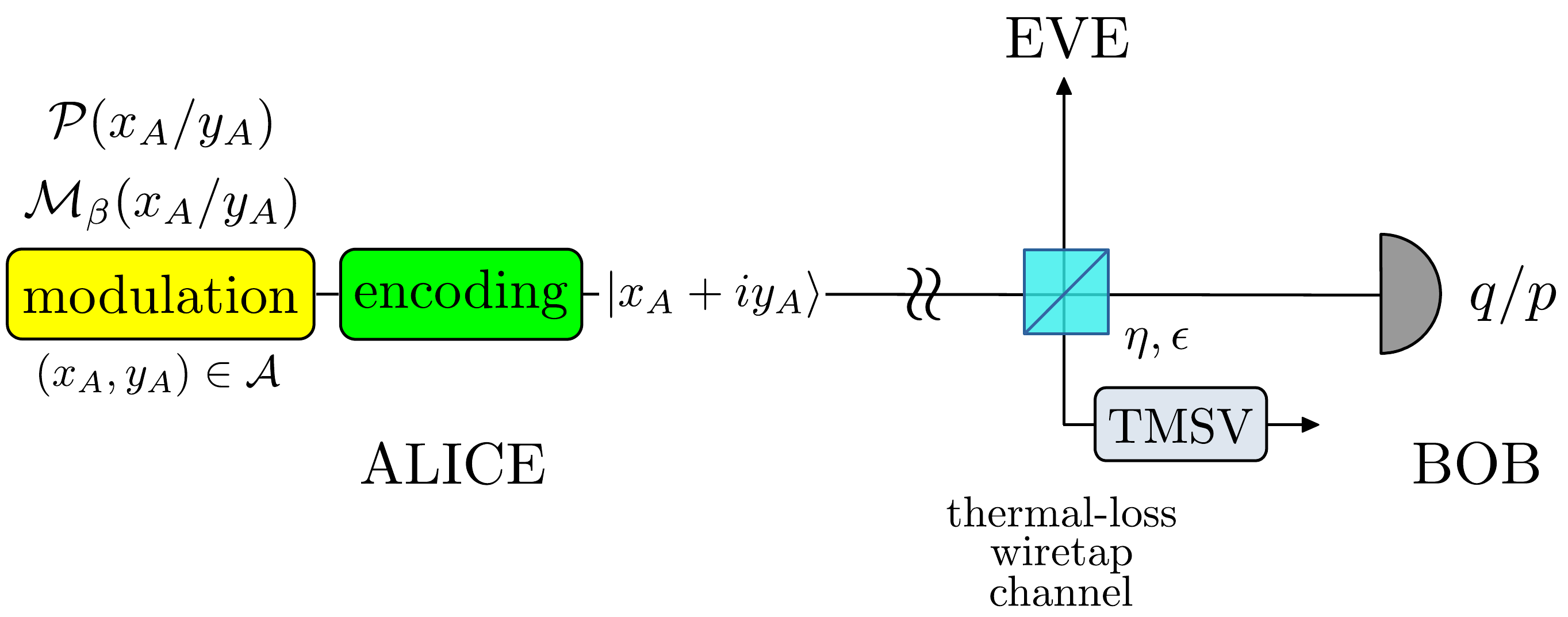} }
\caption{
Scheme of the CV-QKD protocol in the presence of the channel excess noise. Differently from the pure-loss scenario, Eve performs an entangling-cloner attack, that is she injects one arm of a TMSV state into the channel beam splitter, retrieving the final output state.
}\label{fig:07_ExNoiseProto}
\end{figure}

In this section we approach the more realistic scenario that addresses the impact of a thermal channel excess noise on the results of Sec.~\ref{sec: DiscreteProto}. Here the quantum channel is still modelled as a beam splitter with transmissivity $\eta$, but now  a thermal state is injected in the other port. Such a state is characterized by $\bar{n}_{\epsilon} = \eta \epsilon/[2(1-\eta)]$ mean photons, $\epsilon\ge0$  being the introduced excess noise.
In turn, the states probed by Bob are no longer coherent (pure) states but rather displaced thermal (mixed) states \cite{OLIVARES:PLA} and the conditional homodyne probability~(\ref{eq: pB|A}) should be modified accordingly as
\begin{eqnarray}
p_{B|A}(x_{B}\mid x_{A};\epsilon)=\frac{\exp\big[-(x_{B}-2\sigma_{0}\sqrt{\eta}x_{A})^{2}/2\sigma_{\epsilon}^{2}\big]}{\sqrt{2\pi\sigma_{\epsilon}^{2}}}	\label{eq: pB|Aeps}
\end{eqnarray}
with an increased variance $\sigma_{\epsilon}^{2}= \sigma_{0}^{2}(1+\eta \epsilon)$.
Generally speaking, the presence of a channel excess noise is detrimental for CV-QKD, as secure communication only holds up to a maximum transmission distance $d_{\rm max}$, after which the KGR becomes negative \cite{grosshans2005coll, grosshans2007continuous, lodewyck2005controlling, lodewyck2007quantum, e17096072}. The value of $d_{\rm max}$ depends on both the amount of noise and the employed constellation \cite{leverrier2009unconditional, Denys2021explicitasymptotic}.

For the protocol under investigation, the corresponding wiretap channel with non-zero excess noise is depicted in Fig.~\ref{fig:07_ExNoiseProto}, where Eve is assumed to control the added thermal noise by performing an entangling cloner attack \cite{Laudenbach_Rev, Pan2020}. In more detail, she prepares a two-mode squeezed vacuum state (TMSV) with variance $V_\epsilon=1+2\bar{n}_{\epsilon}$ on two modes $\E=(E_1,E_2)$, namely
\begin{eqnarray}
|{\rm TMSV}\rangle\!\rangle= \sqrt{1-\lambda^2} \, \sum_{n=0}^{\infty} \lambda^n \, |n\rangle_{E_1} |n\rangle_{E_2}
\end{eqnarray}
with $\lambda=\sqrt{(V_\epsilon-1)/(V_\epsilon+1)}$ and $|n\rangle$ being the Fock state containing $n$ photons \cite{OLIVARES:PLA}. Thereafter, she injects branch $E_1$ into the channel beam splitter, impinging with the pulse sent by Alice, and, ultimately, collects the output reflected state. Thanks to this strategy, she gets undetected by Alice and Bob, being fully hidden behind the observed excess noise. Indeed, after performing partial trace over modes $\E$, the present scheme is equivalent to a thermal-loss channel with excess noise $\epsilon$ \cite{Laudenbach_Rev}.

\begin{figure}
\centerline{\includegraphics[width=3.0in]{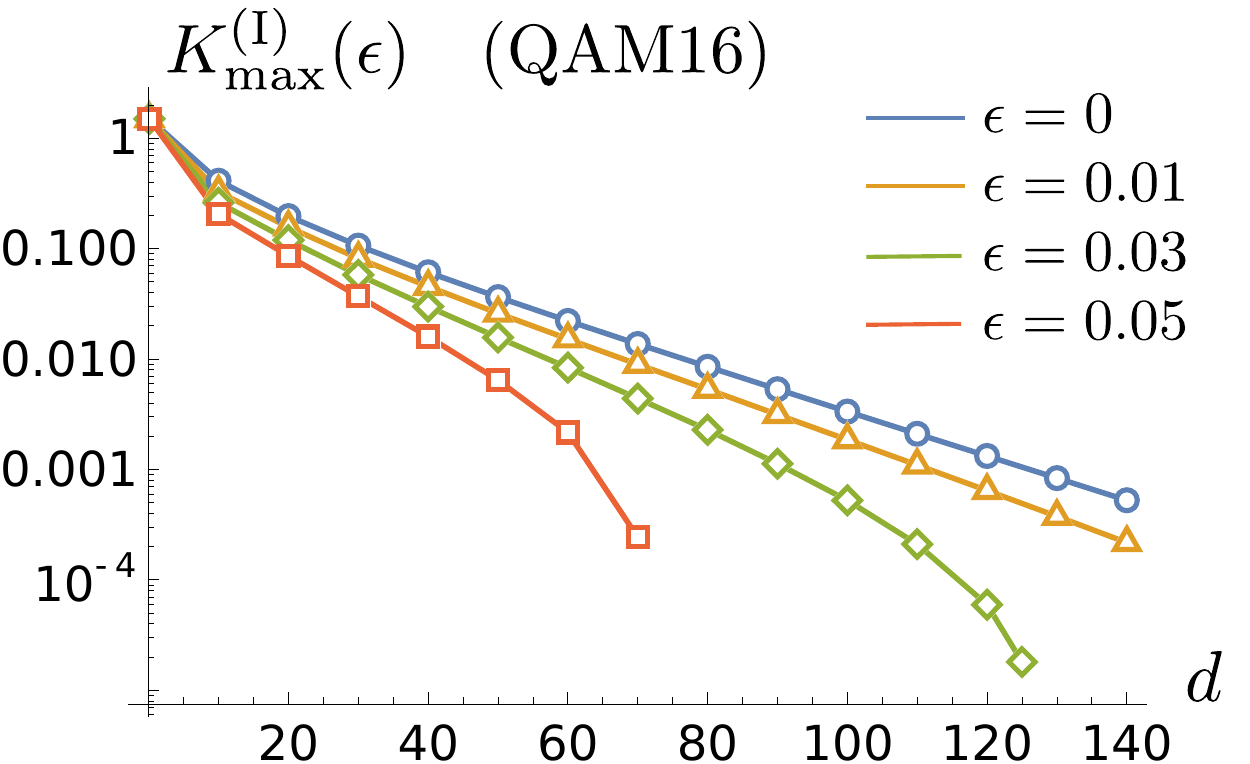} }\vspace*{3ex}
\centerline{\includegraphics[width=3.0in]{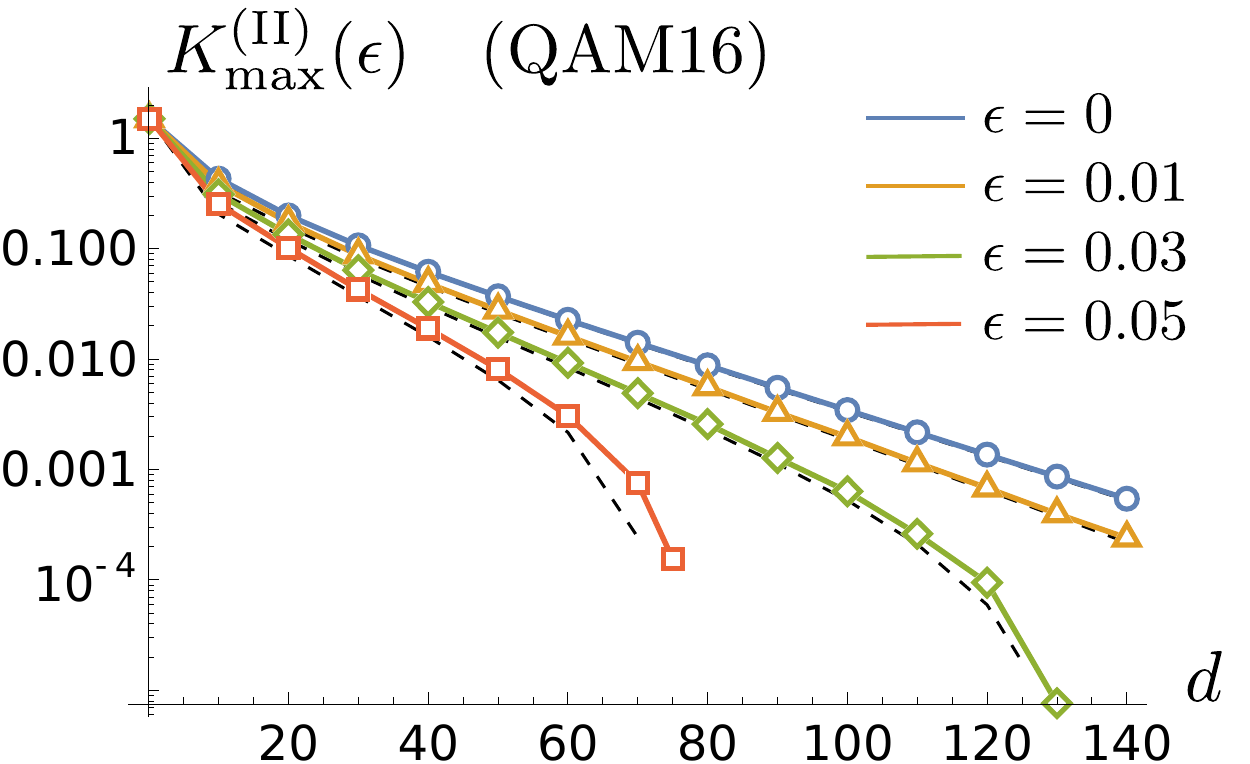} }
\caption{
(Top) Log plot of $K^{(\a)}_{\rm max}(\epsilon)$ as a function of the transmission distance $d$, expressed in km, for different values of the excess noise $\epsilon$. (Bottom) Log plot of $K^{(\b)}_{\rm max}(\epsilon)$ as a function of $d$, for different  $\epsilon$. The black dashed lines are the corresponding KGRs of case $\a$. PAS turns out to be beneficial to increase both the KGR and the maximum transmission distance $d_{\rm max}^{(\p)}$, being more robust against the channel thermal noise. The reconciliation efficiency is $\receff =0.95$.
}\label{fig:08_KGRepsilon}
\end{figure}

As in the previous section, we present the analysis of the KGR addressing only the MB distribution, retrieving cases $\a$ and $\b$ by fixing $\beta=0$ and optimizing $\beta$ with the methods of Sec.~\ref{sec: Optimization}, respectively. 
The mutual information between Alice and Bob reads
\begin{eqnarray}\label{eq: IABthermal}
I_{AB}^{(\p)}(\epsilon) =H_{B}^{(\p)}- \frac12 \log_2 \big(2\pi e \sigma_\epsilon^2\big) \qquad (\p=\a,\b) 
\end{eqnarray}
where $H_{B}^{(\p)}$ is the Shannon entropy of Bob's overall probability distribution 
\begin{eqnarray}
p_{B}^{(\p)}(x_{B};\epsilon)= \sum_{x_A} {\cal M}_\beta (x_A) p_{B|A}(x_{B}\mid x_{A};\epsilon) \,.
\end{eqnarray}
On the contrary, the computation of the Holevo information $\chi_{BE}^{(\p)}(\epsilon)$ is not straightforward and requires the Gaussian formalism \cite{Ferraro2005, Quesada2019}, summarized in App.~\ref{sec: GaussF}.
In particular, if Alice samples the coherent state $|x_A+i y_A\rangle$, both Eve's overall and conditional states $\rho_{\E }(x_A,y_A)$ and $\rho_{\E |x_B}(x_A,y_A)$, respectively, are Gaussian states, whose expressions are derived in App.~\ref{sec: ExNoiseEav}.
In turn, we have
\begin{IEEEeqnarray}{CCL}\label{eq:StatesExN}
\rho_{\E}^{(\p)} &=& \sum_{x_{A},y_{A}}\mathcal{M}_\beta(x_{A})\mathcal{M}_\beta(y_{A})\,
\rho_{\E}(x_A,y_A) \IEEEyessubnumber \\
\rho_{\E |x_{B}}^{(\p)} &=& \frac{1}{p_{B}^{(\p)}(x_{B};\epsilon)}\sum_{x_{A},y_{A}} \mathcal{M}_\beta (x_{A}) \mathcal{M}_\beta (y_{A})\notag\\
& & \hspace{1.0cm} \times p_{B|A}(x_B \mid x_A;\epsilon) \, \rho_{\E |x_B}(x_A,y_A) \, .  \IEEEyessubnumber
\end{IEEEeqnarray}
The Holevo information then reads
\begin{eqnarray}
\chi_{BE}^{(\p)}(\epsilon) =S\big[\rho_{\E }^{(\p)}\big]-\int dx_{B} \, p_B^{(\p)} (x_{B};\epsilon) \, S\big[\rho_{\E |x_{B}}^{(\p)} \big] 
\end{eqnarray}
that can be computed numerically by suitably expanding states~(\ref{eq:StatesExN}) as shown in App.~\ref{sec: GaussF} \cite{Quesada2019}.
The resulting KGR, namely:
\begin{eqnarray}
K^{(\p)}(\epsilon)=\receff  I_{AB}^{(\p)}(\epsilon)-\chi_{BE}^{(\p)}(\epsilon) \qquad (\p=\a,\b) \, ,
\end{eqnarray}
optimized over the modulation energy leads to the maximum achievable rates $K^{(\p)}_{\rm max}(\epsilon)$, $\p=\a,\b$, that are reported in the top and bottom panels of Fig.~\ref{fig:08_KGRepsilon}, respectively, for a QAM16 constellation with different values of $\epsilon$.

As one may expect, the presence of the excess noise results in a lower $K^{(\p)}_{\rm max}(\epsilon)$ with respect to the pure-loss channel one, exhibiting a reduced maximum transmission distance $d_{\rm max}^{(\p)}$. Remarkably, employing PAS is beneficial to increase both the KGR and the value of the maximum distance, as $d_{\rm max}^{(\b)} \ge d_{\rm max}^{(\a)}$: for $\epsilon=0.03$ we have $d_{\rm max}^{(\a)}\approx 125$~km and $d_{\rm max}^{(\b)}\approx 130$~km, while, for $\epsilon=0.05$, $d_{\rm max}^{(\a)}\approx 70$~km and $d_{\rm max}^{(\b)}\approx 75$~km.

\section{A more sophisticated optimization procedure \label{sec: AlternativeOptimiz}}

\begin{figure}
\centerline{\includegraphics[width=3.0in]{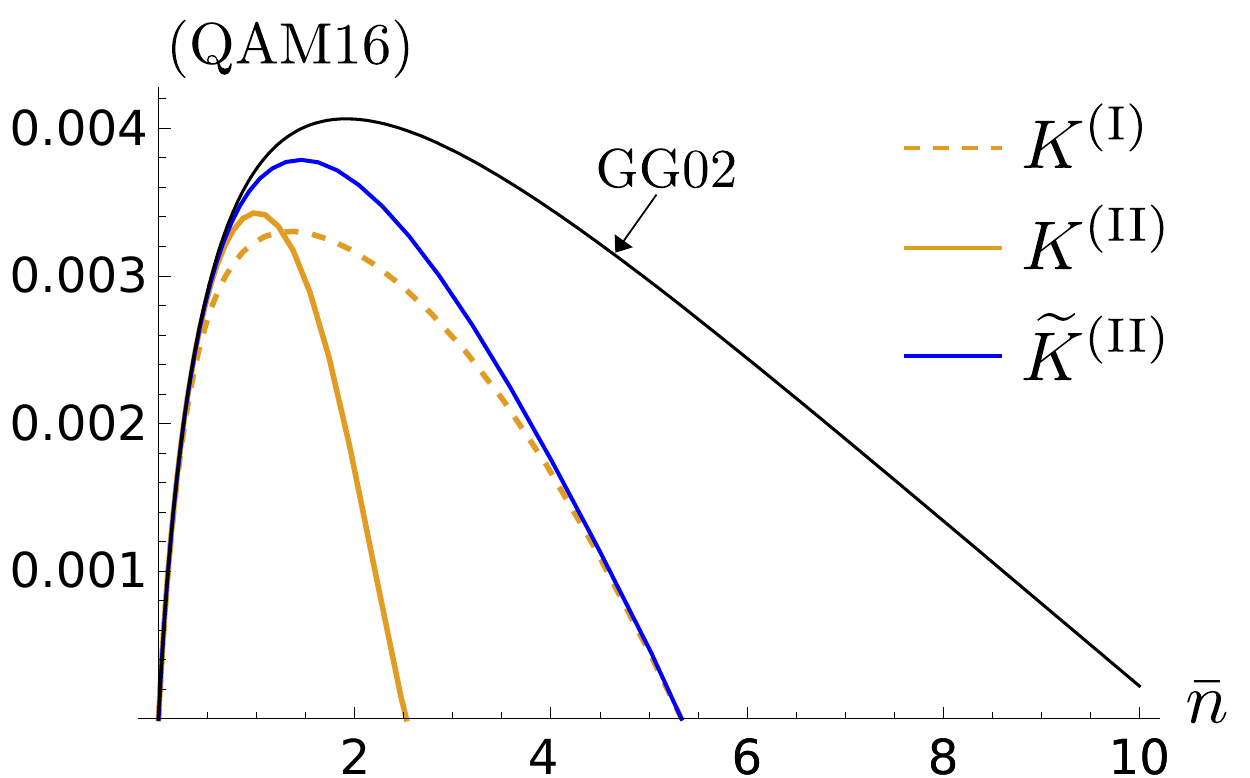} }\vspace*{3ex}
\centerline{\includegraphics[width=3.0in]{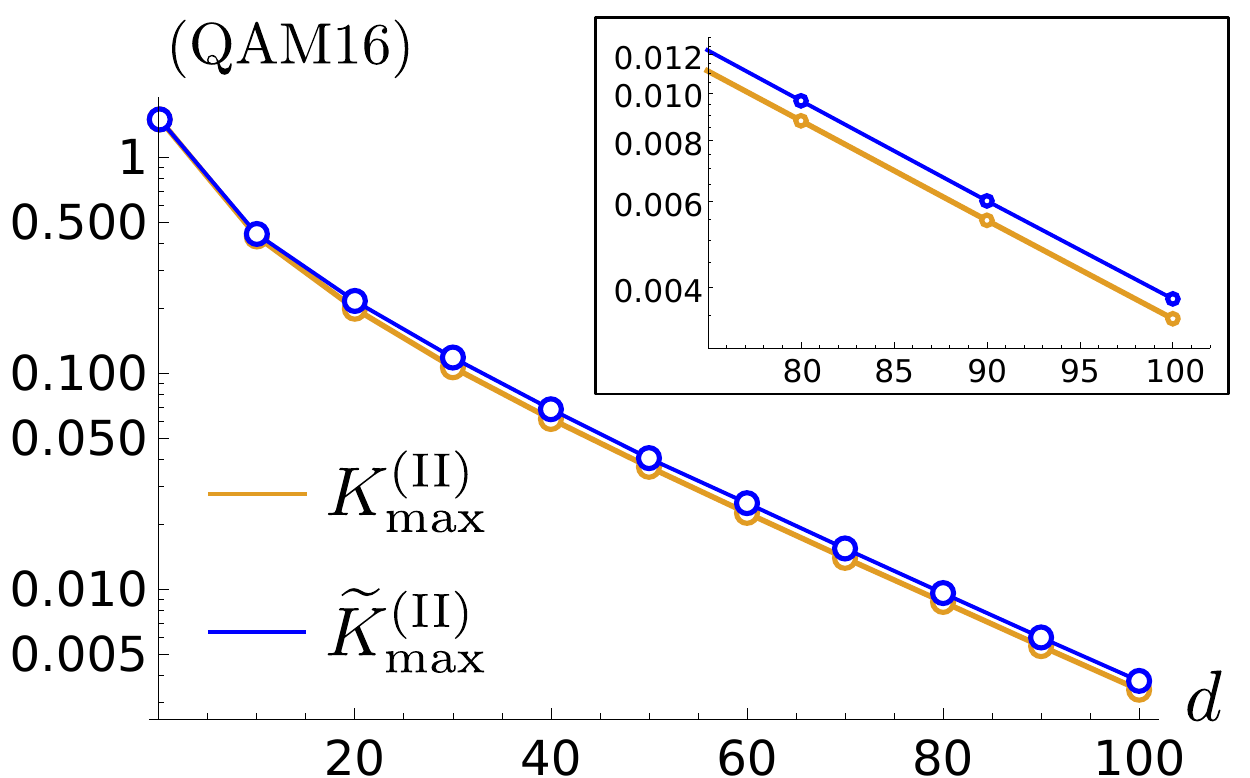} }
\caption{(Top) Plot of $\widetilde{K}^{(\b)}$ and $K^{(\p)}$, $\p=\a,\b$, as a function of the mean energy $\bar{n}$ for $d=100$ km  ($20$~dB attenuation) and a QAM16 ($M=4$) constellation. (Bottom) Log plot of $K^{(\b)}_{\rm max}$ and $\widetilde{K}^{(\b)}_{\rm max}$ for QAM16 as a function of the transmission distance $d$, expressed in km. For $d \gtrsim 80$ km there is an improvement of $\approx 10 \%$ (see the inset). The reconciliation efficiency is $\receff=0.95$.}
\label{fig:09_AlternativeOpt}
\end{figure}

In the previous sections we have addressed case $\b$ by choosing the optimal
values $\beta^{(\b)}$ and $\Delta^{(\b)}$ that maximize the mutual information between Alice and Bob.

However, given the previous discussion, a feasible alternative emerges,
in which the values of the free parameters are selected to maximize
directly the KGR, instead of the sole mutual information $I_{AB}^{(\b)}(\beta)$.
Here we investigate this procedure and, for the sake of simplicity, we only consider the case of a pure-loss channel.

As in the strategy of Sec.~\ref{sec: Optimization}, the spacing $\Delta^{(\b)}(\beta)$ is
obtained as a function of the inverse temperature $\beta$ thanks to the energy constraint
of Eq.~(\ref{eq: FindDeltaB}). However, differently from that strategy, the proper figure of merit becomes the KGR in Eq.~(\ref{eq: Krcoll}), namely,
\begin{eqnarray}
K^{(\b)}(\beta)=\receff I_{AB}^{(\b)}(\beta)-\chi_{BE}^{(\b)} 
\end{eqnarray}
with the quantities $I_{AB}^{(\b)}(\beta)$ and $\chi_{BE}^{(\b)}=\chi_{BE}^{(\b)}(\beta)$ introduced in Eq.s~(\ref{eq: IABb}) and~(\ref{eq: chiBEp}), respectively, and where the dependence on $\beta$ has been highlighted. The optimal inverse temperature is obtained as
\begin{eqnarray}
\widetilde{\beta}^{\, (\b)}=\argmax_{\beta}\, K^{(\b)}(\beta)
\end{eqnarray}
together with the optimal spacing $\widetilde{\Delta}^{(\b)}$ and the optimal KGR
\begin{eqnarray}
\widetilde{K}^{(\b)}=K^{(\b)}\left(\widetilde{\beta}^{\, (\b)}\right) \, .
\end{eqnarray}

Such a task is straightforward but not trivial to implement for the scenario discussed in this paper. Indeed, if we wanted to implement a golden-section-search method as in Sec.~\ref{sec: Optimization},
in any step of the numerical algorithm we should
compute not only $I_{AB}^{(\b)}(\beta)$ but also
the density matrices $\rho_{E}$ and $\rho_{E|x_{B}}$, with
a much higher computational cost. Here, for simplicity, we only 
consider a QAM16 constellation and compare the KGRs $K^{(\p)}$, $\p=\a,\b$, and $\widetilde{K}^{(\b)}$  for distances up to $100$ km.
We report in the top panel of Fig.~\ref{fig:09_AlternativeOpt} the KGRs at $d=100$~km as a function of the modulation energy. As one may expect, we have $\widetilde{K}^{(\b)} \ge K^{(\p)}$. For small $\bar{n}$, $\widetilde{K}^{(\b)}$ (solid blue line) is close to case $\b$ (solid orange line), then it achieves a maximum and decreases approaching the KGR of case $\a$ (dashed orange line).
Remarkably, the present optimization increases further the maximum KGR achievable at a fixed distance. Thus, following the procedure outlined in Sec.~\ref{sec: KGR}, for each distance $d$ we choose the optimal modulation energy and retrieve the maximum achievable value $\widetilde{K}^{(\b)}_{\rm max}$. The comparison between $K^{(\b)}_{\rm max}$ and $\widetilde{K}^{(\b)}_{\rm max}$ in depicted Fig.~\ref{fig:09_AlternativeOpt} (bottom panel). For $d \gtrsim 80$ km there is an improvement of $\approx 10 \%$.
At the same time, these results prove themselves as a validation for
the mutual information optimization method, which is close to the optimization of the
overall key rate, although being a sub-optimal procedure.

\section{Conclusion \label{sec: Concl}}

In this paper we have proposed a CV-QKD protocol employing the discrete QAM 
modulation formats typically exploited in classical telecommunications.
In our proposal, the sender has only a finite set of coherent pulses available,
constituting a square constellation of $M\times M$ states, generated
by sampling either a uniform (case $\a$) or a Maxwell-Boltzmann (case $\b$) distribution. The receiver
performs a homodyne measurement of $q/p$, chosen at random, on his received
signal. For case $\b$, by exploiting PAS, we have evaluated the optimized sender probability distribution
to maximize the shared mutual information.

We have performed the KGR analysis for a pure-loss quantum wiretap channel in reverse reconciliation and compared the obtained KGR with both the associated PSK protocols and the GG02 scheme. We proved QAM modulation as a powerful resource to better approximate the Gaussian modulation and quantified the advantage brought by non-uniform sampling. 
Thereafter, we have addressed the role of a non-zero channel excess noise, whose main detriment is to reduce the KGR, introducing a maximum transmission distance. We compared cases $\a$ and $\b$, showing PAS to be more robust with respect to uniform sampling, allowing to reach larger distances.
Finally, we have proposed a better optimization strategy, based on the direct maximization of the KGR instead of the sole mutual information. We showed this method to outperform both cases $\a$ and $\b$, although case $\b$ may still be considered as a feasible sub-optimal scheme.

We also remark that in our analysis we focused on the asymptotic limit where an infinite dataset is shared between Alice and Bob. In a practical scenario considering finite-size effects, modulation formats that are closer to the continuous modulation may exhibit a lower performance, since there are more terms in the key rate calculation and the convergence can be slower.

Our results prove that a suitable PAS of a discrete constellation allows to overcome the finite
transmitter dynamics and to approximate the performance of the standard CV-QKD based on CM also for increasing average powers. Moreover, they pave the way for the design of feasible schemes compatible with the currently exploited telecom techniques. 

Further improvements may be obtained by improving also the detection stage and investigating the role of non-Gaussian measurements for CV-QKD, such as state-discrimination optimized receivers \cite{cariolaro2015quantum, Eldar2001, Izumi2012, Becerra2013, Chen2018, Notarnicola2022, Notarnicola2023}. Nevertheless, in this case the analysis shall be restricted to a wiretap channel since a security analysis for a fully non-Gaussian protocol is still an open problem.


\appendices
\section{CV-QKD with Gaussian modulation: the GG02 protocol \label{sec: GG02}}
The most relevant protocol of CV-QKD is the so called GG02, proposed by Grosshans and Grangier in 2002 and employing Gaussian modulation \cite{grosshans2002continuous, grosshans2003quantum, grosshans2005coll, grosshans2007continuous}. 
In GG02, Alice encodes information on a continuous ensemble of coherent states $| x_A + i y_A \rangle$, where $x_A, y_A \in \mathbb{R}$ are sampled from the normal distribution $\mathcal{N}_{\Sigma^2}$, namely,
\begin{eqnarray}
{\cal N}_{\Sigma^{2}}(z)=\frac{
\exp\big[
-z^{2}/(2\Sigma^{2})
\big]
}{\sqrt{2\pi\Sigma^{2}}}\,   \qquad(z=x_{A},y_{A})\, .
\end{eqnarray}
Therefore, the overall state generated by Alice reads
\begin{IEEEeqnarray}{CL}
\IEEEyesnumber\label{eq: rho_A CM}
\rho_{A}^{\rm (GG)} &= \int_{\mathbb{R}^2} dx_{A}dy_{A}\, {\cal N}_{\Sigma^2}(x_{A}) { \cal N}_{\Sigma^2}(y_{A})\ket{x_{A}+iy_{A}}\bra{x_{A}+iy_{A}} \nonumber \\[1ex]
&=\nu^{\rm th}(2\Sigma^{2})
\end{IEEEeqnarray}
that is a pseudo-thermal state
\begin{eqnarray}
\nu^{\mathrm{th}}(\bar{n})=
\frac{1}{\bar{n}+1} \sum_{n=0}^{\infty}
\left(
\frac{\bar{n}}{\bar{n}+1} 
\right)^n
| n \rangle \langle n |
\end{eqnarray}
$\bar{n}=2\Sigma^{2}$ being the mean energy per symbol and $| n \rangle$ representing the Fock state with $n$ photons \cite{OLIVARES:PLA}. 

For the scheme investigated in this paper, that is a pure-loss channel followed by homodyne detection by Bob, the mutual information reads \cite{grosshans2002continuous, grosshans2005coll, cover06}
\begin{eqnarray}
I_{AB}^{\rm (GG)}= \frac12 \log_{2}(1+2 \eta \bar{n})\label{eq: I_AB CM}
\end{eqnarray}
which coincides with the capacity of
an AWGN channel as derived in the Shannon-Hartley 
theorem \cite{Shannon48, Gallager68, taub1986principles,d1995information}.

By performing the security analysis in a reverse reconciliation scenario, if Eve exploits
an entangling cloner attack, the Holevo information reads \cite{grosshans2002continuous, grosshans2003quantum, grosshans2007continuous}
\begin{IEEEeqnarray}{CCL}
\IEEEyesnumber
\chi_{BE}^{\rm (GG)} &=& \log_{2}\bigg(\frac{V_{E}+1}{\bar{V}_{E}+1}\bigg)+\frac{V_{E}-1}{2}\log_{2}\bigg(\frac{V_{E}+1}{V_{E}-1}\bigg)
\notag\\[1ex]
& & \hspace{2.5cm}
 -\frac{\bar{V}_{E}-1}{2}\log_{2}\bigg(\frac{\bar{V}_{E}+1}{\bar{V}_{E}-1}\bigg) 	\label{eq: chi_BE CM} 
\end{IEEEeqnarray}
where
\begin{IEEEeqnarray}{CCL}
\IEEEyesnumber
V_{E} &=&1+2(1-\eta)\bar{n} \IEEEyessubnumber \\[1ex]
\bar{V}_{E} &=&\sqrt{\frac{\eta+(1-\eta)(1+2\bar{n})}{1-\eta+\eta(1+2\bar{n})}(1+2\bar{n})}	\, . \IEEEyessubnumber
\end{IEEEeqnarray}
The resulting KGR writes
\begin{eqnarray}\label{eq: K GG02}
K^{\rm (GG)}=\receff  I_{AB}^{\rm (GG)}-\chi_{BE}^{\rm (GG)} \, .
\end{eqnarray}
Finally, as discussed in the main text, for each transmission distance it is possible to optimize Eq.~(\ref{eq: K GG02}) over the modulation energy, obtaining the maximum achievable KGR $K^{\rm (GG)}_{\rm max}$ and the maximum energy $\bar{n}^{\rm (GG)}_{\rm max}$.

\section{A brief outline of the Gaussian formalism} \label{sec: GaussF}
The Gaussian state formalism is a useful tool to perform the KGR analysis in the presence of a thermal-loss channel. Here we present a brief outline of its main features \cite{Quesada2019, Ferraro2005}. 
We consider a $n$-mode optical field, described by the bosonic field operators $a_k$, $k=1,\ldots,n$, satisfying the canonical commutation relations $[a_k,a_l]=0$, $[a_k,a_l^{\dagger}]=\delta_{kl}$, and their corresponding quadrature operators
\begin{eqnarray}
    q_k= \sigma_0 (a_k+ a_k^{\dagger}) \quad \mbox{and} \quad p_k= i \sigma_0 (a_k^{\dagger}-a_k)
\end{eqnarray}
such that $[q_k,p_l]=2 i \sigma_0^2 \delta_{kl}$, $\sigma_0^2$ being the shot-noise variance. In the following we will exploit a vector notation by introducing the operator $\hat{\mathbf{r}}= (q_1, p_1, q_2, p_2, \ldots, q_n, p_n)^\mathsf{T}$. 

\subsection{Gaussian states}

A quantum state $\rho$ is a Gaussian state if its associated Wigner function is Gaussian, namely
\begin{eqnarray}\label{eq:Wcart}
   W[\rho](\mathbf{r}) =  \frac{1}{(2\pi)^n \sqrt{\det(\boldsymbol{\sigma})}} \,  \exp \left[ 
   - \frac12 (\mathbf{r}-\mathbf{x})^\mathsf{T} \, \boldsymbol{\sigma}^{-1} \, (\mathbf{r}-\mathbf{x})
   \right]
\end{eqnarray}
where $\mathbf{r}^\mathsf{T}= (x_1, y_1, \ldots ,x_n, y_n) \in \mathbb{R}^{2n}$ and
\begin{IEEEeqnarray}{CCL}
    \mathbf{x}&=& {\rm Tr}[\rho \, \hat{\mathbf{r}}] \IEEEyessubnumber \\
    \boldsymbol{\sigma} &=& \frac12  {\rm Tr} \left[ \rho \, \{(\hat{\mathbf{r}}-\mathbf{x}),(\hat{\mathbf{r}}-\mathbf{x})^\mathsf{T}\} \right] \IEEEyessubnumber
\end{IEEEeqnarray}
are the first moment vector (FM) and the covariance matrix (CM), $\{A,B\}=AB+BA$ being the anti-commutator of $A$ and $B$ \cite{Ferraro2005}.
In turn, a Gaussian state is characterized by its FM and CM. Eq.~(\ref{eq:Wcart}) may also be re-expressed as
\begin{eqnarray}\label{eq:Wcomplex}
   W[\rho](\boldsymbol{\alpha}) =\frac{1}{\pi^n \sqrt{\det(\widetilde{\boldsymbol{\sigma}})}} \,  \exp \left[ 
   - \frac12 (\boldsymbol{\alpha}-\boldsymbol{\beta})^{\dagger} \,\, \widetilde{\boldsymbol{\sigma}}^{-1} \, (\boldsymbol{\alpha}-\boldsymbol{\beta})
   \right]
\end{eqnarray}
with $\boldsymbol{\alpha}^\mathsf{T}= (\alpha_1, \alpha^*_1, \ldots ,\alpha_n, \alpha^*_n) \in \mathbb{C}^{2n}$, and
\begin{eqnarray}
\boldsymbol{\beta}= U \mathbf{x} \quad \mbox{and} \quad \widetilde{\boldsymbol{\sigma}} = U \boldsymbol{\sigma} U^{\dagger}
\end{eqnarray}
where $U= \oplus_{k=1}^{n} U_1$ and
\begin{eqnarray}
U_1=
\frac{1}{2\sigma_0} \, 
\begin{pmatrix}
1 & i \\
1 & -i 
\end{pmatrix}
 \, .
\end{eqnarray}

The expansion of a Gaussian state $\rho$ onto the Fock basis has been recently derived in \cite{Quesada2019}. To this aim, we introduce the matrices
\begin{IEEEeqnarray}{CCL}
\boldsymbol{\sigma}_Q&=& \widetilde{\boldsymbol{\sigma}} + \mathbb{I}_{2n}/2  \IEEEyessubnumber\\
\boldsymbol{A}&=& \boldsymbol{X} (\mathbb{I}_{2n}-\boldsymbol{\sigma}_Q^{-1}) \IEEEyessubnumber \\
\boldsymbol{\gamma}^{\mathsf T} &=& \boldsymbol{\beta}^{\dagger} \boldsymbol{\sigma}_Q^{-1} \IEEEyessubnumber
\end{IEEEeqnarray}
where $\mathbb{I}_{2n}$ is the $2n\times 2n$ identity matrix and $\boldsymbol{X}= \oplus_{s=1}^n \boldsymbol{\sigma}_x$, $\boldsymbol{\sigma}_x$ being the Pauli $x$-matrix.
Then, the matrix element $\rho_{\boldsymbol{mk}}= \langle \boldsymbol{m}|\rho |\boldsymbol{k}\rangle$, $|\boldsymbol{k}\rangle=|k_1 \, k_2 \, \ldots k_n\rangle$ and $|\boldsymbol{m}\rangle=|m_1 \, m_2 \, \ldots m_n\rangle$, reads
\begin{eqnarray}
\rho_{\boldsymbol{mk}} = T_{\boldsymbol{mk}}  \, \prod_{s=1}^{n} \partial_{\alpha_s}^{k_s} \partial_{\alpha^*_s}^{m_s} \,
 \exp \left(\frac12 \boldsymbol{\alpha}^{\mathsf T} \boldsymbol{A} \boldsymbol{\alpha} + \boldsymbol{\gamma}^{\mathsf T} \boldsymbol{\alpha}\right)\Big|_{\boldsymbol{\alpha}=0}
\end{eqnarray}
where
\begin{eqnarray}
T_{\boldsymbol{mk}} =\frac{1}{\sqrt{\det(\boldsymbol{\sigma}_Q) \prod_{s=1}^{n} k_s! m_s!}}\exp \left(-\frac12 \boldsymbol{\beta}^{\dagger} \boldsymbol{\sigma}_Q^{-1} \boldsymbol{\beta}\right) \, .
\end{eqnarray}

\subsection{Gaussian dynamics}

Gaussian dynamics is provided by unitary evolution generated by bilinear Hamiltonians and is associated with a symplectic matrix $S$. Given an input Gaussian state $\rho_{\rm in}$ with $({\bf x}_{\rm in}, \boldsymbol{\sigma}_{\rm in})$, the evolved state $\rho_{\rm out}$ is still Gaussian with FM and CM given by \cite{Ferraro2005}
\begin{eqnarray}
   {\bf x}_{\rm out} = S \, {\bf x}_{\rm in} \quad \mbox{and} \quad
    \boldsymbol{\sigma}_{\rm out} = S \, \boldsymbol{\sigma}_{\rm in} \, S^\mathsf{T} \,
\end{eqnarray}
respectively.

Finally, we discuss the case of conditional dynamics \cite{Ferraro2005}. We consider a bipartite system $AE$, composed of $n_{A}$ an $n_E$ optical modes, respectively, prepared in a Gaussian state $\rho_{AE}$ with $\mathbf{x}=(\mathbf{x}_A,\mathbf{x}_E)$ and CM (written in block form)
\begin{eqnarray}\label{eq:Block}
\boldsymbol{\sigma} = 
\begin{pmatrix}
\boldsymbol{\sigma}_A & \boldsymbol{\sigma}_{Z} \\ 
\boldsymbol{\sigma}_{Z}^\mathsf{T} & \boldsymbol{\sigma}_E
\end{pmatrix}
\, .
\end{eqnarray}
Thereafter, a Gaussian measurement is performed on subsystem $A$, associated with the CM $\boldsymbol{\sigma}_m$, retrieving the outcome $\mathbf{x}_m \in \mathbb{R}^{2n_A}$. The resulting conditional state $\rho_{E|\mathbf{r}_m}$ on modes $E$ is still a Gaussian state with FM and CM equal to
\begin{IEEEeqnarray}{CCL}\label{eq:CondDyn}
    \mathbf{x}_{E|\mathbf{x}_m} &=& \mathbf{x}_E + \boldsymbol{\sigma}_{Z}^{\mathsf T} \left(\boldsymbol{\sigma}_A+\boldsymbol{\sigma}_m \right)^{-1}\left(\mathbf{x}_m-\mathbf{x}_A \right) \IEEEyessubnumber \\
    \boldsymbol{\sigma}_{E|\mathbf{r}_m} &=& \boldsymbol{\sigma}_E - \boldsymbol{\sigma}_{Z}^\mathsf{T} \left(\boldsymbol{\sigma}_A+\boldsymbol{\sigma}_m \right)^{-1} \boldsymbol{\sigma}_{Z}
    \IEEEyessubnumber
\end{IEEEeqnarray}
respectively.

\section{Calculation of Eve's states for a thermal-loss wiretap channel}\label{sec: ExNoiseEav}

Here we derive explicitly the quantum states $\rho_{\E }(x_A,y_A)$ and $\rho_{\E |x_B}(x_A,y_A)$ introduced in Sec.~\ref{sec: ExNoise}. The entire computation is based on the Gaussian formalism outlined in the previous appendix.

In the presence of a thermal-loss wiretap channel, Eve generates a TMSV state with variance $V_\epsilon=1+2\bar{n}_\epsilon$, $\bar{n}_\epsilon=\eta \epsilon/[2(1-\eta)]$, with zero FM, ${\bf x}_{\E }^{(0)} =0$, and the CM
\begin{eqnarray}
\boldsymbol{\sigma}_{\E }^{(0)} = \sigma_0^2 \,
\begin{pmatrix}
V_\epsilon \, \mathbb{I}_2 & Z_\epsilon \, \boldsymbol{\sigma}_z \\ 
Z_\epsilon\,  \boldsymbol{\sigma}_z  & V_\epsilon \, \mathbb{I}_2
\end{pmatrix}
\end{eqnarray}
with $Z_\epsilon=\sqrt{V_\epsilon^2-1}$ and $\boldsymbol{\sigma}_z$ being the Pauli $z$-matrix.
Moreover, if Alice samples the coherent state $|x_A+i y_A\rangle$, she gets a single-mode Gaussian state with FM ${\bf x}_A^{(0)}= 2\sigma_0 \,(x_A,y_A)$ and CM $\boldsymbol{\sigma}_A^{(0)}=\sigma_0^2 \, \mathbb{I}_2$.
Thereafter, Alice's pulse interferes at the channel beam splitter with Eve's mode $E_1$, resulting in a tripartite Gaussian state $\rho_{A\E }(x_A,y_A)$ characterized by FM and CM 
equal to
\begin{IEEEeqnarray}{CCL}
{\bf x}_{A\E } &=& S \, \left({\bf x}_A^{(0)} \oplus {\bf x}_{\E }^{(0)}\right) \IEEEyessubnumber \\
\boldsymbol{\sigma}_{A\E } &=& S \, \left(\boldsymbol{\sigma}_A^{(0)} \oplus \boldsymbol{\sigma}_{\E }^{(0)}\right) \, S^\mathsf{T} \IEEEyessubnumber
\end{IEEEeqnarray}
with $S= S_{\rm BS} \oplus \mathbb{I}_2$ and
\begin{eqnarray}
S_{\rm BS} =
\begin{pmatrix}
\sqrt{\eta} \, \mathbb{I}_{2} & \sqrt{1-\eta} \, \mathbb{I}_{2} \\
-\sqrt{1-\eta} \, \mathbb{I}_{2} & \sqrt{\eta} \, \mathbb{I}_{2}
\end{pmatrix}
\end{eqnarray}
being the symplectic matrix associated with the beam splitter operation \cite{Ferraro2005}.

In turn, Eve's overall state $\rho_{\E }(x_A,y_A)$ is characterized by $({\bf x}_{\E }, \boldsymbol{\sigma}_{\E })$, whilst her conditional state after Bob's homodyne measurement $\rho_{\E |x_B}(x_A,y_A)$ is associated with FM ${\bf x}_{\E |x_B}$ and $\boldsymbol{\sigma}_{\E |x_B}$ equal to Eq.~(\ref{eq:CondDyn}),
with $\mathbf{x}_m=(x_B,0)$ and
\begin{eqnarray}
\boldsymbol{\sigma}_m =
\lim_{z\rightarrow 0} \, 
\begin{pmatrix}
z & 0 \\
0 & 1/z
\end{pmatrix}
\end{eqnarray}
being the CM associated with homodyne detection \cite{Ferraro2005}.



\ifCLASSOPTIONcaptionsoff
  \newpage
\fi




\begin{thebibliography}{10}
\baselineskip 12pt
\providecommand{\url}[1]{#1}
\csname url@samestyle\endcsname
\providecommand{\newblock}{\relax}
\providecommand{\bibinfo}[2]{#2}
\providecommand{\BIBentrySTDinterwordspacing}{\spaceskip=0pt\relax}
\providecommand{\BIBentryALTinterwordstretchfactor}{4}
\providecommand{\BIBentryALTinterwordspacing}{\spaceskip=\fontdimen2\font plus
\BIBentryALTinterwordstretchfactor\fontdimen3\font minus
  \fontdimen4\font\relax}
\providecommand{\BIBforeignlanguage}[2]{{%
\expandafter\ifx\csname l@#1\endcsname\relax
\typeout{** WARNING: IEEEtran.bst: No hyphenation pattern has been}%
\typeout{** loaded for the language `#1'. Using the pattern for}%
\typeout{** the default language instead.}%
\else
\language=\csname l@#1\endcsname
\fi
#2}}
\providecommand{\BIBdecl}{\relax}
\BIBdecl

\bibitem{RSA:rivest1978}
R.~L. Rivest, A.~Shamir, and L.~Adleman, ``A method for obtaining digital
  signatures and public-key cryptosystems,'' \emph{Commun. ACM}, vol.~21,
  no.~2, pp.~120--126, 1978.

\bibitem{vernam1926cipher}
G.~S. Vernam, ``Cipher printing telegraph systems for secret wire and radio
  telegraphic communications,'' \emph{Trans. AIEE}, vol.~XLV, pp.~295--301,
  1926.

\bibitem{Shannon49}
C.~E. Shannon, ``Communication theory of secrecy systems,'' \emph{Bell Syst.
  Tech. J.}, vol.~28, no.~4, pp.~656--715, 1949.

\bibitem{BB84:bennet1984}
C.~H. Bennett and G.~Brassard, ``Quantum cryptography: Public-key distribution
  and coin tossing,'' in \emph{Proc. {IEEE} Int. Conf. on Computers, Systems
  and Signal Processing}, Bangalore, India, 1984, pp.~175--179.

\bibitem{ekert1992quantum}
A.~K. Ekert, ``Quantum cryptography and {B}ell’s theorem,'' in \emph{Quantum
  Measurements in Optics}.\hskip 1em plus 0.5em minus 0.4em\relax Springer,
  1992, pp.~413--418.

\bibitem{grosshans2002continuous}
F.~Grosshans and P.~Grangier, ``Continuous variable quantum cryptography using
  coherent states,'' \emph{Phys. Rev. Lett.}, vol.~88, Art.~no.~057902, Jan 2002.

\bibitem{gisin2002quantum}
N.~Gisin, G.~Ribordy, W.~Tittel, and H.~Zbinden, ``Quantum cryptography,''
  \emph{Rev. Mod. Phys.}, vol.~74, pp.~145--195, Mar 2002.

\bibitem{wootters1982single}
W.~K. Wootters and W.~H. Zurek, ``A single quantum cannot be cloned,''
  \emph{Nature}, vol.~299, no.~5886, pp.~802--803, 1982.

\bibitem{Ralph}
T.~C. Ralph, ``Continuous variable quantum cryptography,'' \emph{Phys. Rev. A},
  vol.~61, Art.~no.~010303, Dec 1999.

\bibitem{grosshans2003quantum}
 F.~Grosshans {\em et al.},
  ``Quantum key distribution using {G}aussian-modulated coherent states,''
  \emph{Nature}, vol.~421, no.~6920, pp.~238--241, 2003.

\bibitem{grosshans2005coll}
F.~Grosshans, ``Collective attacks and unconditional security in continuous
  variable quantum key distribution,'' \emph{Phys. Rev. Lett.}, vol.~94, Art.~no.~020504, Jan 2005.

\bibitem{grosshans2007continuous}
F.~Grosshans, A.~Ac{\'\i}n, and N.~Cerf, ``Continuous-variable quantum key
  distribution,'' in \emph{Quantum information with continuous variables of
  atoms and light}.\hskip 1em plus 0.5em minus 0.4em\relax World Scientific,
  2007, pp.~63--83.

\bibitem{e17096072}
E.~Diamanti and A.~Leverrier, ``Distributing secret keys with quantum
  continuous variables: Principle, security and implementations,''
  \emph{Entropy}, vol.~17, no.~9, pp.~6072--6092, 2015.

\bibitem{OLIVARES:PLA}
S.~Olivares, ``Introduction to generation, manipulation and characterization of
  optical quantum states,'' \emph{Phys. Lett. A}, vol.~418, Art.~no.~127720, Dec
  2021.

\bibitem{lodewyck2005controlling}
J.~Lodewyck, T.~Debuisschert, R.~Tualle-Brouri, and P.~Grangier, ``Controlling
  excess noise in fiber-optics continuous-variable quantum key distribution,''
  \emph{Phys. Rev. A}, vol.~72, Art.~no.~050303, Nov 2005.

\bibitem{lodewyck2007quantum}
J.~Lodewyck {\em et al.}, 
``Quantum key distribution over 25 km with an all-fiber
  continuous-variable system,'' \emph{Phys. Rev. A}, vol.~76, Art.~no.~042305, Oct
  2007.

\bibitem{fossier2009field}
S.~Fossier {\em et al.}, 
``Field test of a continuous-variable quantum key distribution
  prototype,'' \emph{New J. Phys.}, vol.~11, no.~4, Art.~no.~045023, Apr 2009.

\bibitem{hillery2000quantum}
M.~Hillery, ``Quantum cryptography with squeezed states,'' \emph{Phys. Rev. A},
  vol.~61, Art.~no.~022309, Jan 2000.

\bibitem{reid2000quantum}
M.~D. Reid, ``Quantum cryptography with a predetermined key, using
  continuous-variable {E}instein-{P}odolsky-{R}osen correlations,'' \emph{Phys.
  Rev. A}, vol.~62, Art.~no.~062308, Nov 2000.

\bibitem{Shannon48}
C.~E. Shannon, ``A mathematical theory of communication,'' \emph{Bell Syst.
  Tech. J.}, vol.~27, no.~3/4, pp.~379--423/623--656, July/Oct 1948.

\bibitem{Gallager68}
R.~G. Gallager, \emph{Information Theory and Reliable Communication}.\hskip 1em
  plus 0.5em minus 0.4em\relax New York: Wiley, 1968.

\bibitem{Jouguet2012}
P.~Jouguet, S.~Kunz-Jacques, E.~Diamanti, and A.~Leverrier, ``Analysis of imperfections in practical continuous-variable quantum key distribution," \emph{Phys. Rev. A}, vol.~86, Art.no.~032309, Sep. 2012.

\bibitem{DjordjevicDGM}
I.~B.~Djordjevic, ``On the Discretized Gaussian Modulation (DGM)- Based Continuous Variable-QKD," \emph{IEEE Access}, vol.~7, pp.~65342-65346, May 2019.

\bibitem{leverrier2009unconditional}
A.~Leverrier and P.~Grangier, ``Unconditional security proof of long-distance
  continuous-variable quantum key distribution with discrete modulation,''
  \emph{Phys. Rev. Lett.}, vol.~102, Art.~no.~180504, May 2009.

\bibitem{leverrier2010continuous}
A.~Leverrier and P.~Grangier, ``Continuous-variable quantum key distribution protocols with a discrete modulation,'' arXiv:1002.4083~[quant-ph], 2010.

\bibitem{leverrier2011continuous}
A.~Leverrier and P.~Grangier, ``Continuous-variable quantum-key-distribution protocols with a non-{G}aussian modulation,'' \emph{Phys. Rev. A}, vol.~83, Art.~no.~042312, Apr
  2011.

\bibitem{becir2012continuous}
A.~Becir, F.~A.~A. El-Orany, and M.~R.~B. Wahiddin, ``Continuous-variable
  quantum key distribution protocols with eight-state discrete modulation,''
  \emph{Int. J. Quantum Inf.}, vol.~10, no.~01, Art.~no.~1250004, 2012.

\bibitem{hirano2017implementation}
T.~Hirano {\em et al.}, 
``Implementation of continuous-variable
  quantum key distribution with discrete modulation,'' \emph{Quantum Sci.
  Technol.}, vol.~2, no.~2, Art.~no.~024010, 2017.

\bibitem{Qu2017}
Z.~Qu, and I.~B.~Djordjevic, ``Four-Dimensionally Multiplexed Eight-State Continuous-Variable Quantum Key Distribution Over Turbulent Channels," \emph{IEEE Phot. J.}, vol.~9, no.~6, pp.~1-8, Dec. 2017.

\bibitem{ghorai2019asymptotic}
S.~Ghorai, P.~Grangier, E.~Diamanti, and A.~Leverrier, ``Asymptotic security of
  continuous-variable quantum key distribution with a discrete modulation,''
  \emph{Phys. Rev. X}, vol.~9, Art.~no.~021059, Jun 2019.

\bibitem{lin2019asymptotic}
J.~Lin, T.~Upadhyaya, and N.~L\"utkenhaus, ``Asymptotic security analysis of
  discrete-modulated continuous-variable quantum key distribution,''
  \emph{Phys. Rev. X}, vol.~9, Art.~no.~041064, Dec 2019.

\bibitem{Liao2020}
Q.~Liao, G.~Xiao, C.-G.~Xu, Y.~Xu, and Y.~Guo, ``Discretely modulated continuous-variable quantum key distribution with an untrusted entanglement source," \emph{Phys. Rev. A}, vol.~102, Art.no.~032604, Sep. 2020.

\bibitem{Ghalaii2020}
M.~Ghalaii {\em et al}, ``Discrete-Modulation Continuous-Variable Quantum Key Distribution Enhanced by Quantum Scissors," \emph{IEEE J. on Sel. Areas in Commun.}, vol.~38, no.~3, pp.~506-516, Mar. 2020.

\bibitem{Denys2021explicitasymptotic}
A.~Denys, P.~Brown, and A.~Leverrier, ``Explicit asymptotic secret key rate of
  continuous-variable quantum key distribution with an arbitrary modulation,''
  \emph{{Quantum}}, vol.~5, Art.~no.~540, Sep. 2021.

\bibitem{Papanastasiou2021}
P.~Papanastasiou, and S.~Pirandola, ``Continuous-variable quantum cryptography with discrete alphabets: Composable security under collective Gaussian attacks," \emph{Phys. Rev. Research}, vol.~3, Art.no.~013047, Jan. 2021.

\bibitem{Kanitschar2022}
F.~Kanitschar, and C.~Pacher, ``Optimizing Continuous-Variable Quantum Key Distribution with Phase-Shift Keying Modulation and Postselection," \emph{Phys. Rev. Applied}, vol.~18, Art.~no.~034073, Sep. 2022.

\bibitem{Djordjevic2019}
I.~B.~Djordjevic, ``Optimized-Eight-State CV-QKD Protocol Outperforming Gaussian Modulation Based Protocols," \emph{IEEE Phot. J.}, vol.~11, no.~4, pp.~1-10, Aug. 2019.

\bibitem{Almeida2021}
M.~Almeida {\em et al.}, ``Secret key rate of multi-ring M-APSK continuous variable quantum key distribution," \emph{Opt. Express}, vol.~29, no.~23, pp.~38669-38682, Nov. 2021. 

\bibitem{Pereira2022}
D.~ Pereira{\em et al.}, ``Probabilistic shaped 128-APSK CV-QKD transmission system over optical fibres," \emph{Opt. Lett.}, vol.~47, no.~15, pp.~3948-3951, Aug. 2022.

\bibitem{Roumestan2021}
F.~ Roumestan {\em et al.}, ``High-rate continuous variable quantum key distribution based on probabilistically shaped 64 and 256-QAM," in \emph{2021 European Conference on Optical Communication (ECOC)}. IEEE, 2021, pp.~1--4.

\bibitem{Roumestan2022}
F.~ Roumestan {\em et al.}, ``Experimental Demonstration of Discrete Modulation Formats for Continuous Variable Quantum Key Distribution," arXiv:2207.11702 [quant-ph], 2022.

\bibitem{bocherer2015bandwidth}
G.~B{\"o}cherer, F.~Steiner, and P.~Schulte, ``Bandwidth efficient and
  rate-matched low-density parity-check coded modulation,'' \emph{IEEE Trans.
  Commun.}, vol.~63, no.~12, pp.~4651--4665, 2015.

\bibitem{Buchali:JLT2016}
F.~Buchali {\em et al.}, 
  ``Rate adaptation and reach increase by probabilistically shaped 64-{QAM}: An
  experimental demonstration,'' \emph{J. Light. Technol.}, vol.~34, no.~7, pp.~1599--1609, 2016.

\bibitem{fehenberger2016probabilistic}
T.~Fehenberger, A.~Alvarado, G.~B{\"o}cherer, and N.~Hanik, ``On probabilistic
  shaping of quadrature amplitude modulation for the nonlinear fiber channel,''
  \emph{J. Light. Technol.}, vol.~34, no.~21, pp.~5063--5073, 2016.

\bibitem{Cai2004}
N.~Cai, A.~Winter, and R.~W.~Yeun, ``Quantum privacy and quantum wiretap channels," \emph{Probl. Inf. Transm.}, vol.~40, no.~4, pp.~318-336, Oct. 2004.

\bibitem{Pan2020}
Z.~Pan {\em et al.}, ``Secret-Key Distillation across a Quantum Wiretap Channel under Restricted Eavesdropping," \emph{Phys. Rev. Applied}, vol.~14, Art.~no.~024044, Aug. 2020.

\bibitem{CCDM:schulte2016}
P.~Schulte and G.~B{\"o}cherer, ``Constant composition distribution matching,''
  \emph{IEEE Trans. Inf. Theory}, vol.~62, no.~1, pp.~430--434, 2015.

\bibitem{HIDM:civelli2020}
S.~Civelli and M.~Secondini, ``Hierarchical distribution matching for
  probabilistic amplitude shaping,'' \emph{Entropy}, vol.~22, no.~9, Art.~no.~958,
  2020.

\bibitem{ESS:gultekin2020}
Y.~C. G{\"u}ltekin, T.~Fehenberger, A.~Alvarado, and F.~M. Willems,
  ``Probabilistic shaping for finite blocklengths: Distribution matching and
  sphere shaping,'' \emph{Entropy}, vol.~22, no.~5, Art.~no.~581, 2020.

\bibitem{kschischang1993optimal}
F.~R. Kschischang and S.~Pasupathy, ``Optimal nonuniform signaling for
  {G}aussian channels,'' \emph{IEEE Trans. Inf. Theory}, vol.~39, no.~3, pp.~913--929, 1993.

\bibitem{buchali2015experimental}
F.~Buchali {\em et al.}, 
  ``Experimental demonstration of capacity increase and rate-adaptation by
  probabilistically shaped 64-{QAM},'' in \emph{2015 European Conference on
  Optical Communication (ECOC)}.\hskip 1em plus 0.5em minus 0.4em\relax IEEE,
  2015, pp.~1--3.

\bibitem{fehenberger2016sensitivity}
T.~Fehenberger {\em et al.}, 
  ``Sensitivity gains by mismatched probabilistic shaping for optical
  communication systems,'' \emph{IEEE Photon. Technol. Lett.}, vol.~28, no.~7,
  pp.~786--789, 2016.

\bibitem{Banaszek2021}
K.~Banaszek, M.~Jachura, P~Kolenderski and M.~Lasota, 
``Optimization of intensity-modulation/direct-detection optical key distribution under passive eavesdropping,"
{\em Opt. Express}, vol.~29, pp.~43091--43103, 2021.

\bibitem{proakis01}
J.~G. Proakis, \emph{Digital Communications}, 4th~ed.\hskip 1em plus 0.5em
  minus 0.4em\relax McGraw Hill, 2001.

\bibitem{cariolaro2015quantum}
G.~Cariolaro, \emph{Quantum communications}.\hskip 1em plus 0.5em minus
  0.4em\relax Springer, 2015.

\bibitem{van2004reconciliation}
G.~Van~Assche, J.~Cardinal, and N.~J. Cerf, ``Reconciliation of a
  quantum-distributed {G}aussian key,'' \emph{IEEE Trans. Inf. Theory},
  vol.~50, no.~2, pp.~394--400, 2004.

\bibitem{bennett1995generalized}
C.~H. Bennett, G.~Brassard, C.~Cr{\'e}peau, and U.~M. Maurer, ``Generalized
  privacy amplification,'' \emph{IEEE Trans. Inf. Theory}, vol.~41, no.~6, pp.~1915--1923, 1995.

\bibitem{agrawal02}
G.~P. Agrawal, \emph{Fiber-optic communications systems}, 3rd~ed.\hskip 1em
  plus 0.5em minus 0.4em\relax Wiley, 2002.

\bibitem{Jouguet_Kunz-Jacques_Leverrier_Grangier_Diamanti_2013}
P.~Jouguet {\em et al.}, 
  ``Experimental demonstration of long-distance continuous-variable quantum key
  distribution,'' \emph{Nat. Photonics}, vol.~7, no.~5, pp.~378–381, May 2013.

\bibitem{Lodewyck_Debuisschert_Tualle-Brouri_Grangier_2005}
J.~Lodewyck, T.~Debuisschert, R.~Tualle-Brouri, and P.~Grangier, ``Controlling
  excess noise in fiber-optics continuous-variable quantum key distribution,''
  \emph{Phys. Rev. A}, vol.~72, no.~5, Art.~no.~050303, Nov 2005.

\bibitem{olivares:PhaseSpace}
S.~Olivares, ``Quantum optics in the phase space,'' \emph{Eur. Phys. J. Spec.
  Top.}, vol.~203, no.~1, pp.~3--24, Apr 2012.

\bibitem{cover06}
T.~M. Cover and J.~A. Thomas, \emph{Elements of Information Theory},
  2nd~ed.\hskip 1em plus 0.5em minus 0.4em\relax Hoboken, NJ: Wiley, 2006.

\bibitem{GoldenSection}
W.~H. Press, S.~A. Teukolsky, W.~T. Vetterling, and B.~P. Flannery,
  \emph{Numerical recipes 3rd edition: The art of scientific computing}.\hskip
  1em plus 0.5em minus 0.4em\relax Cambridge University Press, 2007.

\bibitem{GaussOpt1}
M.~Navascu\'es, F.~Grosshans, and A.~Ac\'{\i}n, ``Optimality of Gaussian Attacks in Continuous-Variable Quantum Cryptography," {\em Phys. Rev. Lett.}, vol.~97, Art.~no.~190502, Nov. 2006.

\bibitem{GaussOpt2}
A.~Leverrier, ``Theoretical study of continuous-variable quantum key distribution," PhD Thesis, {T{\'e}l{\'e}com ParisTech}, 2009.

\bibitem{GaussOpt3}
R.~Garc\'{\i}a-Patr\'on and N.~J. Cerf, ``Unconditional Optimality of Gaussian Attacks against Continuous-Variable Quantum Key Distribution," {\em Phys. Rev. Lett.}, vol.~97, Art.~no.~190503, Nov. 2006.

\bibitem{Laudenbach_Rev}
F.~Laudenbach {\em et al.}, ``Continuous‐variable quantum key distribution with Gaussian modulation—the theory of practical implementations," {\em Adv. Quantum Technol.}, vol.~1, no.~1, p.~1800011, Jun. 2018.

\bibitem{holevo1998capacity}
A.~S. Holevo, ``The capacity of the quantum channel with general signal
  states,'' \emph{IEEE Trans. Inf. Theory}, vol.~44, no.~1, pp.~269--273, 1998.

\bibitem{Ferraro2005}
A. Ferraro, S. Olivares and M.~G.~A. Paris,
{\em Gaussian States in quantum information}, Bibliopolis Napoli, 2005.

\bibitem{Quesada2019}
N.~Quesada {\em et al.}, ``Simulating realistic non-Gaussian state preparation,"
\emph{Phys. Rev. A}, vol.~100, Art.~no.~022341, Aug. 2019.

\bibitem{Eldar2001}
Y.~C.~Eldar, ``On Quantum Detection and the Square-Root Measurement," \emph{IEEE Trans. Inf. Theory}, vol.~47, no.~3, pp.~858-872, Mar. 2001.

\bibitem{Izumi2012}
S.~Izumi {\em et al.}, ``Displacement receiver for phase-shift-keyed coherent states," \emph{Phys. Rev. A}, vol.~86, Art.~no.~042328, Oct. 2012.

\bibitem{Becerra2013}
F.~E.~Becerra {\em et al.}, ``Experimental demonstration of a receiver beating the standard quantum limit for multiple nonorthogonal state discrimination," \emph{Nature Photon.}, vol.~7, pp.~147–152, Jan. 2013.

\bibitem{Chen2018}
T.~Chen, K.~Li, Y.~Zuo, and B.~Zhu, ``Hybrid quantum receiver for quadrature amplitude modulation coherent-state discrimination beating the classical limit," \emph{Appl. Opt.}, vol.~57, pp.~817-822, Feb. 2018.

\bibitem{Notarnicola2022}
M.~N.~Notarnicola, M.~G.~A.~Paris, and S.~Olivares, ``Hybrid near-optimum binary receiver with realistic photon-number-resolving detectors," \emph{J. Opt. Soc. Am. B}, vol.~40, pp.~705-714, Mar. 2023. 

\bibitem{Notarnicola2023}
M.~N.~Notarnicola, M.~Jarzyna, S.~Olivares, and K.~Banaszek, ``Optimizing state-discrimination receivers for continuous-variable quantum key distribution over a wiretap channel," arXiv:2306.11493 [quant-ph], 2023.

\bibitem{taub1986principles}
H.~Taub and D.~L. Schilling, \emph{Principles of communication systems}.\hskip
  1em plus 0.5em minus 0.4em\relax McGraw-Hill Higher Education, 1986.

\bibitem{d1995information}
G.~D’Ariano, C.~Macchiavello, and M.~G.~A.~Paris, ``Information gain in quantum
  communication channels,'' in \emph{Quantum Communications and
  Measurement}.\hskip 1em plus 0.5em minus 0.4em\relax Springer, 1995, pp.~339--350.

\end{thebibliography}


%




\end{document}